\documentclass[11pt]{article}
\usepackage{authblk}
\usepackage{a4wide}
\usepackage{url}
\usepackage{graphicx}
\usepackage{dcolumn}
\usepackage{bm}
\usepackage{xcolor}
\usepackage[utf8]{inputenc}
\usepackage{mathptmx}
\usepackage{ulem}
\usepackage{amsfonts}
\UseRawInputEncoding

\title{Fostering Innovation: Streamlining Magnetocaloric Materials Research by Digitalization}

\author[1]{Simon Bekemeier}
\author[2]{Moritz Blum}
\author[3,4]{Luana Caron}
\author[5]{Alisa Chirkova}
\author[2]{Philipp Cimiano}
\author[2]{Basil Ell*}
\author[3]{Inga Ennen}
\author[5]{Michael Feige}
\author[3]{Maik Gaerner}
\author[5]{Thomas Hilbig}
\author[3]{Andreas H\"utten}
\author[3]{G\"unter Reiss}
\author[3]{Tapas Samanta}
\author[5]{Sonja Sch\"oning}
\author[3,5]{Christian Schr\"oder*}
\author[5]{Lennart Schwan}
\author[3]{Chris Taake}
\author[5]{Martin Wortmann}

\affil[1]{Federal Institute for Materials Research and Testing, 12205 Berlin, Germany}
\affil[2]{Faculty of Technology and CITEC, Bielefeld University, 33619 Bielefeld, Germany}
\affil[3]{Faculty of Physics, Bielefeld University, 33615 Bielefeld, Germany}
\affil[4]{Helmholtz-Zentrum Berlin for Materials and Energy, 12489 Berlin, Germany}
\affil[5]{Bielefeld Institute for Applied Materials Research, Bielefeld University of Applied Sciences and Arts, 33619 Bielefeld, Germany}
\begin{document}

\maketitle

*Corresponding Authors, bell@techfak.uni-bielefeld.de, christian.schroeder@hsbi.de
\thispagestyle{empty}

\begin{abstract}
Refrigeration based on the magnetocaloric effect (MCE) can contribute to energy-saving, environmentally friendly cooling in private households, or industrial application. The cooling is based on the reversible heat release or uptake during a phase-transformation of the materials that can be controlled by a magnetic field. This process could replace conventional compression-based refrigeration, which often relies on environmentally harmful refrigerants. Here we show, how to digitalize the process chain for the synthesis, theoretical and experimental characterization, and prototypical application of magnetocaloric alloy. Different Heusler alloys are examined experimentally as model systems for potential application in magnetic cooling. OTTR templates are used for the acquisition and semantic representation of knowledge in the development of an ontology. The ontology, when combined with unstructured data, can be exploited to train a model that can then be used to predict missing facts, which can help to gain new insights and to generate new hypotheses. Furthermore, tools are developed that automate data acquisition into ontological structures and workflows are implemented that provide an easy-to-use theoretical and experimental evaluation of the MCE from first principles and raw data. 
\end{abstract}


\section{Introduction}
The magnetocaloric effect (MCE) is the result of the strong coupling between magnetic and lattice degrees of freedom and manifests as changes in temperature and entropy in response to an applied magnetic field change. Magnetic materials have been used in a “one-shot” process to achieve sub-Kelvin temperatures since the late 1930’s \cite{Giauque1933, Debye1926} but it was only half a century later that the first applications using a regenerative cycle for cooling around room-temperature was proposed \cite{Brown1976}. The MCE is naturally maximal around phase transitions and is observed in a large number of magnetic alloys and compounds. Thus, research on MCE materials involves the screening and optimization of many different materials systems, generating a large amount of data that, so far, has not been organized in a systematic manner, hindering the progress of the field. 

Beyond the perspective of physics, we consider a broader perspective. Research projects contribute to the advancement of scientific knowledge. The obtained knowledge exists in the minds of the researchers involved and is, to some extent, reflected in the data being generated such as machine configuration files and files that contain measurement results. It is well understood that knowledge and data need to be captured in some form and be managed as a basis for knowledge sharing, exploitation, and future advancement of knowledge. Obviously, knowledge representation techniques need to be employed for capturing knowledge in digital form. Here, we develop and use an ontology and make use of existing ontologies for the purpose of managing knowledge and data relevant to and resulting from our research activities.

According to Studer \textit{et al.} \cite{Studer1998}, an ``ontology is a formal, explicit specification of a shared conceptualization.'' In short, creating an ontology for a domain of interest (e. g., chemistry) encompasses a group of people agreeing on how to define the scope of the domain, how to conceptualize that domain, and which terms to use to describe the elements of this conceptualization. More specifically, agreement needs to be reached about which things are relevant in that domain (e. g., chemical elements, chemical molecules, acids) and which terms to use to refer to these things (e. g., define that the term “salt” refers to the chemical substance NaCl). Furthermore, agreement needs to be reached about the relevant attributes of the things relevant in that domain (e. g., the number of valence electrons of a chemical element) and which terms to use to refer to these attributes (e. g., the term that refers to the number of valence electrons is ``valence''). Also, agreement needs to be reached about relations between these things relevant in the domain and which terms to use to refer to these relations (e. g., one would like to express the fact that every full-Heusler compound is a Heusler compound and decide that the term that refers to this relation is ``is\_a\_type\_of'').
Moreover, agreement needs to be reached about the attributes of the relations (e. g. ``is\_a\_type\_of'' is a transitive relation, which means that if A is a type of B and if B is a type of C, then it is true that A is a type of C).

To create an ontology means to express the aforementioned agreements using an ontology language such as RDF (Resource Description Framework), RDFS (Resource Description Framework Schema), or OWL (Web Ontology Language). The result is a knowledge graph that can then, for example, be stored in a database commonly referred to as a triple store and can then be queried using the query language SPARQL. Using a knowledge graph instead of a relational database offers several benefits. One key advantage is that a knowledge graph can provide a semantic representation of data, which means that via logical entailment one can derive statements from a knowledge graph via logical deduction, which means that from true statements one always derive true statements. Those derived statements are \textit{implied} by the knowledge graph, but the knowledge graph does not have to explicitly contain these statements. In contrast, from a relational database one does not entail statements. A knowledge graph has semantics (i. e., it can express more than what it explicitly contains) if it makes use of symbols that have a formally defined semantics, such as terms from the RDFS vocabulary. These terms have a semantics because they occur in entailment rules, which specify how from statements that contain these terms one can derive potentially new statements. Terms that have a formally defined semantics can be used to formally define the semantics of other symbols.

An example of a symbol from the RDFS vocabulary with formally defined semantics is rdfs:subClassOf which can, for example, be used to express that every full-Heusler compound is a type of Heusler compound (e. g., via the statement “full-Heusler rdfs:subClassOf Heusler”) and that a Heusler compound is a type of compound (e. g., via the statement “Heusler rdfs:subClassOf compound”). The RDFS vocabulary contains an entailment rule that formally defines that rdfs:subClassOf is a transitive relation. This rule can be described as follows. Given a statement that expresses that something (let’s call it A) is in the relation rdfs:subClassOf with something else (let’s call it B), and a statement that expresses that B is in the relation rdfs:subClassOf with something else (let’s call it C), then the rule entails a statement that expresses that A is in the relation rdfs:subClassOf with C. Thereby, via the rule it is formally described that the relation rdfs:subClassOf is a transitive relation. Coming back to our example, this rule allows us to derive the fact that every full-Heusler is a compound.

Aside from the elementary illustration concerning the formal expression of transitivity in a relation, the ontology language OWL facilitates the articulation of more intricate statements. For instance, it can be used to assert that a molecule containing a carboxylic group is a carboxylic acid, that AlmondAllergy is synonymous with an allergy triggered by almonds \cite{Horrocks2013}, and that a carbon atom does not form more than four stable covalent bonds. For a comprehensive introduction to knowledge graphs, we refer to Hogan \textit{et al.} \cite{Hogan2021};  for an introduction to ontologies we refer to Uschold and Gruninger \cite{Uschold1996};  for a discussion about the role of ontologies in knowledge management we refer to Abecker and van Elst \cite{Abecker2009}. 

In the laboratory, manual data transfer methods are often employed, which can result in repetitive tasks, inconsistent data handling and inefficient collaboration. To address these challenges, the we introduce an automated data acquisition pipeline that standardizes data handling and integrates experimental metadata into a structured format aligned with the established ontology. By integrating Python Jupyter notebooks with an electronic laboratory notebook, e.g. eLabFTW \cite{CARPi2017} in our case, the pipeline ensures reproducibility and seamless data transfer for further analysis, allowing researchers to focus on scientific innovation rather than mundane data management. We have implemented two branches, one computational and one experimental, both starting with a compound description and ultimately resulting in the evaluation of the entropy change $\Delta S$, a key metric for the performance of magnetocaloric materials. The computational branch uses tools based on the Density Functional Theory (DFT) and an in-house Markov Chain Monte Carlo (MCMC) spin dynamics code to simulate magnetization curves. The experimental branch automates data acquisition from synthesized samples. Both branches yield magnetization curves enabling further analysis of magnetic phase transitions. 

\section{Structure of the Ontology}
Knowledge representation poses multiple challenges, especially in developing large ontologies. These challenges include making multiple design decisions at the same time and collaborating with domain experts unfamiliar with the technologies behind ontologies, like RDF or OWL. An ontology engineering methodology based on Reasonable Ontology Templates (OTTR) \cite{skjaeveland2024} can help overcome these issues \cite{skjaeveland2018, blum2023}. OTTR introduces an abstraction layer over the backbone semantic web technologies behind ontologies, thereby facilitating communication with domain experts. Moreover, OTTR decouples the process of determining \textit{what} to model from \textit{how} to model it, enabling a more agile and iterative development process. Lastly, an OTTR template-based ontology engineering methodology favors a bottom-up (from data to ontology) engineering approach, as the existing structured data can be utilized in designing the OTTR templates. The final ontology, called DiProMag ontology, is publicly available \cite{dipromagonotology}. We provide the T-Box, as well as all templates and example template instances. The documentation of all templates is automatically generated from our OTTR template library and can be found at \cite{dipromag_onto}. All data can also be found on Zenodo \cite{zenodo}. An overview over the PMD core ontology and the ontologies created by the other projects
can be found in Bayerlein \textit{et al.}~\cite{Bayerlein2025}.

\begin{figure}
  \includegraphics[width=\linewidth]{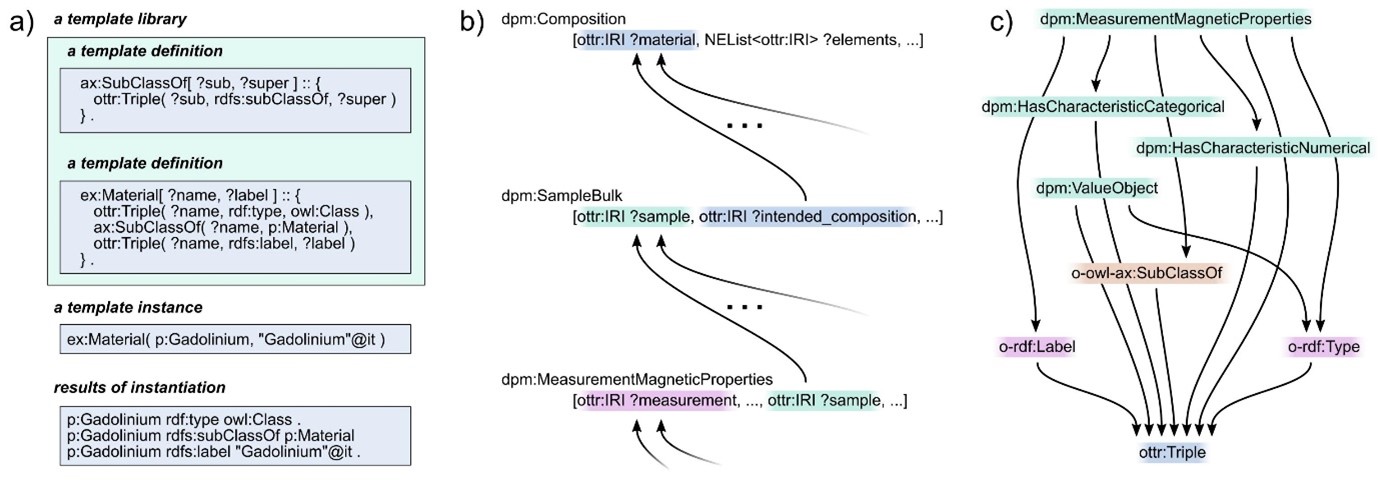}
  \caption{a) an example template library and an instance of a template. The terms p:Gadolinium and ''Gadolinium''@it are bound to the template parameters ?name and ?label, respectively, to create an instance of the template, whose triples are shown underneath. b) Relations between the example templates. c) Template call hierarchy of the dpm:MeasurementMagneticProperties template.}
  \label{fig1}
\end{figure}

When developing an ontology with OTTR, the ontology is usually represented by an OTTR template library and a set of OTTR template instances, from which an RDF serialization such as a file in Turtle syntax can be generated, if needed. An OTTR template library is a set of OTTR template definitions. See \textbf{Figure \ref{fig1}a} for an example. Each template definition consists of a header (e. g., ``\texttt{ax:SubClassOf{[}?sub, ?super{]}}'' --- here, \texttt{ax:SubClassOf} is the template\textquotesingle s name; and \texttt{?sub} and \texttt{?super} are its parameters --- and a body (enclosed by curly brackets). Template bodies can refer to other templates --- thus, templates can be nested. \texttt{ottr:Triple} is the base template that creates RDF triples. A template instantiation denotes binding concrete values to the template parameters. Instantiating a template leads to RDF triples to be created.

In this paper we distinguish between \textit{template instantiation}, i.e., the process of instantiating a template, and \textit{template instance}, i. e., the result of instantiating a template. Furthermore, we distinguish between templates that are only instantiated from within
other templates (such as \texttt{ax:SubClassOf} in Figure \ref{fig1}a), and those that are not (such as \texttt{pz:Pizza} in Figure \ref{fig1}). The latter we call user-facing templates. User-facing templates are instantiated by domain experts, or these template instances are generated by some tool.
We developed an ontology engineering methodology that is based on OTTR templates. The subsequent steps of the methodology are: 1) define the scope of the ontology and screen the available data, 2) design of template headers, 3) template header design verification and
documentation, 4) design of template bodies, 5) handling of axiomatic triples, 6) template body documentation, 7) template library documentation, and 8) template instantiation \& data integration. It is important to note that this list of steps does not follow a strictly linear process model. In the following we introduce the steps in detail, report our practical experience, and present our template library, which we have developed.
\subsection{OTTR-centric Ontology Engineering Methodology}
Here, we sketch the main steps of the methodology that has been described in \cite{blum2023}.
\subsubsection{Define the scope of the ontology and screen the available
data}
We define the ontology\textquotesingle s scope through competency questions (CQs). In general, our CQs are not designed specifically for our OTTR template-based ontology engineering methodology, and, therefore, do not differ from CQs developed in the context of other ontology engineering methodologies. Within our framework, these competency questions ask for knowledge that would also be published in research papers, like results of experiments, but also details required for the reproduction of experiments.

Example competency questions we developed are:
\begin{itemize}
\item
  What is the stoichiometry of a considered sample? E. g., sample x has stoichiometry Mn\textsubscript{1.82}Cu\textsubscript{0.18}Sb.
\item
  Which steps are performed in the production of a compound? E. g., first melting at temperature x, secondly ...
\item
  What are the outcomes of a measurement using method x? E. g., MPMS provides $\Delta T$ hysteresis and maximum $\Delta S$ values.
\end{itemize}

In parallel to the definition of CQs, we started to collect example data holding the information to answer the CQs. During data collection, we noticed a substantial degree of heterogeneity in the data management practices applied to the data. To facilitate future data integration, we not only collected representative data samples but also metadata like
where the data is located (i. e., machine, volume, folder, etc.), how it can be accessed (i. e., file, database, service), how often it is updated, and the format of the files.
\subsubsection{Design of template headers}
Based on the specified scope of the ontology and the collected data, OTTR template headers can be developed. An OTTR template header mainly specifies a template by giving it a name and listing its parameter names together with their data types. When designing template headers, trade-offs need to be made, taking the principles like complexity of templates vs. complexity of template relations into account. For a complete list of important design principles and further details, we refer to \cite{blum2023}.

As potential template candidates, we used one template per process step, one template per material type, one template per synthesis method, and one template per measurement method. The logical relation between those steps served as a basis for connecting the templates
(later explained and denoted as template relations). Next, we derived the template parameters directly from the available data. In cases where there is no data available to answer a certain CQ, the domain experts provided us with some artificially created data similar to the expected real data. The resulting template headers were iteratively improved to follow the design principles. Some example template headers we developed are:

\begin{itemize}
\item
  \textbf{dpm:Composition}{[}~ottr:IRI ?material,
  NEList\textless ottr:IRI\textgreater{} ?elements,
  NEList\textless xsd:float\textgreater{} \\
  ?stoichiometry\_portion\_unit\_at\_percent~{]}
\item
  \textbf{dpm:SampleBulk}{[}\\ottr:IRI ?sample, \\
  xsd:string
  ?sample\_code,\\
  ? ottr:IRI ?intended\_compound,\\
  NEList\textless owl:Class\textgreater{} ?elements,\\
  NEList\textless xsd:float\textgreater{} ?purities\_unit\_percent,\\
  NEList\textless owl:Class\textgreater{} ?shapes,\\
  NEList\textless xsd:float\textgreater{} ?target\_masses\_unit\_kg,\\
  NEList\textless xsd:float\textgreater{} ?actual\_masses\_unit\_kg,\\
  NEList\textless xsd:float\textgreater{} ?mass\_deviations\_unit\_kg,\\
  ? xsd:float ?overall\_weight\_unit\_kg,\\
  ? ottr:IRI ?production\_process~{]}
\item
  \textbf{dpm:MeasurementMagneticProperties}{[}~\\
  ottr:IRI ?measurementMagneticPropertiesInitCharacteristicsMounting,\\
  ? owl:Class ?machine,\\
  ? owl:Class ?magnetic\_moment\_component,\\
  ottr:IRI ?sample,\\
  ? xsd:float ?mass\_unit\_kg,\\
  ? xsd:float ?volume\_mass\_unit\_m3,\\
  ? owl:Class ?shape, \\
  ? owl:Class ?sample\_orientation,\\ xsd:string
  ?sample\_orientation\_additional\_description,\\
  ? owl:Class ?holder, \\
  xsd:float ?force\_unit\_N="0"\^{}\^{}xsd:float,\\
  ? xsd:int ?field\_cycles\_number\_unit\_number,\\
  ? xsd:boolean ?zero\_field\_cooled~{]}
\end{itemize}

For example, the dpm:Composition template creates the ontological representation for a given composition when instantiated. The template takes three parameters: the ?material parameter gets the IRI (Internationalized Resource Identifier) of the composition to allow other IRIs and templates to refer to this composition, e. g. ex:Ni2MnGa; the ?elements parameter gets a list of IRIs of chemical elements the composition is made of, e. g., ex:Ni, ex:Mn, ex:Ga; and the ?stoichiometry\_portion\_unit\_at\_percent parameter gets a list of float values representing the portion of the elements in atomic percent, e. g., 1, 2, 1, 1 for the considered composition.

The dpm:SampleBulk template allows to describe bulk samples and the dpm:MeasurementMagneticProperties template allows to describe the input and configuration of a magnetic properties measurement. Detailed explanations of the usage of these templates can be found in the documentation.

If a template (header) has many parameters, then it can become difficult to use or difficult to update the template (header). Thus, it can be advisable to split a template header into multiple template headers while at the same time not distributing parameters that form a group over multiple templates. Consider the situation where we model a pizza with OTTR templates. Either we create one entity for the pizza that links to all the pizza\textquotesingle s properties, or we create an entity for the pizza that links to both the entities of the dough and the toppings, where dough and topping are described separately. In the former case, all information about the pizza would be passed to a single template. In the latter case, one would first instantiate the topping template and the dough template, and then instantiate the pizza template where the identifiers of the dough and the topping are passed as parameter values. However, distributing information across multiple templates can increase complexity for end users, as they need to understand how these templates are related so that they can be used correctly. Formally, if a template is split into multiple templates, the templates definitions are not related. Only if the templates are instantiated, the instances are related. 

Example template relations connecting the templates dpm:Composition, dpm:SampleBulk, and dpm:MeasurementMagneticProperties are shown in Figure \ref{fig1}b. For example, the IRI bound to the parameter ?material in the template dpm:Composition is the same as the IRI bound to the parameter ?intended\_composition in the template dpm:SampleBulk. Thereby, a relation of instances between these two templates is established. 

We distinguish template relations from the template call hierarchy, i.e., which templates are used inside of other templates. As an example, Figure \ref{fig1}c shows the template call hierarchy of the dpm:MeasurementMagneticProperties template.

\subsubsection{Template Header Design Verification and Documentation}
Verifying template headers is crucial at this stage to avoid issues in the development of template bodies. The verification ensures that all relevant information can be captured, there are no parameters for irrelevant pieces of information, and there are no redundant parameters.
Further, it ensures that instantiations of multiple templates are bound together reasonably if specific information is distributed across multiple templates.

Using our OTTR Extension \cite{mediawiki} for Semantic MediaWiki (SMW), domain experts can validate user-facing template headers by filling out forms with sample data. This extension allows for the definition and instantiation of OTTR templates directly within SMW, generating user-friendly input forms for template headers
automatically. During the validation process for the ontology, minor issues with the template headers were identified, such as that a parameter was defined as a single value where instead a list of values was necessary.

To simplify template instantiation with multiple relationships, we propose defining workflows that describe how templates are related. Figure \ref{fig1}b shows relations between the example templates. These workflows, developed based on template dependencies, ensure a connected
ontology. For instance, a workflow describes that materials should be specified conceptually before their measured properties are modeled. Additionally, the documentation of user-facing templates includes a general description, information about modeling limitations, and
specific details such as data types, parameter names, default values, and examples. Documenting parameters in tables (parameter name, data type, default values, example values, and description) ensures easy understanding, usability, and correct modeling of domain knowledge
during template body development.

\subsubsection{Design of Template Bodies}
In the previous steps, the template headers were developed, which define the shape of the input data. In this step, template bodies are developed for these template headers by ontology engineers, based on the template's signatures and the documentation. As proposed by other ontology engineering methodologies, we encourage the reuse of existing ontologies with approved modeling decisions inside the template bodies to ensure interoperability with other ontologies. The reuse of existing ontologies and following common standards makes the ontology interoperable with other ontologies and easily queryable with SPARQL. The ontology engineers make use of the template header documentation to identify relevant ontologies and ontology design patterns (ODPs) for re-use. In detail, we make use of CCO \cite{cco} (Common Core Ontologies), PMDco \cite{pmdco} (PMD core ontologies) and QUDT \cite{qudt} (Quantities, Units, Dimensions, and Types Ontology) in the development of our ontology.

The development of the template bodies is straightforward, as a pre-structuring is done through developing template definitions and through the ontologies that will be reused. Here, we refer to established ontology engineering methodologies that can be used to develop the ontology inside the template bodies. The use of sub-templates leads to a faster development process, as similar data elements occur frequently across our data and can be covered by the same sub-templates. We observed that a collection of similar physical experiments often share the same data about the environment where the experiments take place, like temperature and humidity. This hints at the fact that a sub-template should be introduced to model the (shared) data about the environment. It is important to note that domain experts do not need to understand the template bodies. Instead, we propose to show them graph visualizations of the template relations and seek the domain experts' feedback on the sufficiency of the graphs\textquotesingle{} contents. The main requirement is that the ontology can be queried to answer the competency questions, which can be done in the ontology validation step.

\subsubsection{Inclusion of Axiomatic Triples into the Ontology}
If a template body makes use of a term (i. e., the template body mentions it) about which we want to specify axioms, then the template\textquotesingle s body instantiates another template that encapsulates all axiomatic triples about that term. Thus, that template is the only place where the axiomatic triples are stored, and an ontology engineer knows where to find these axioms. E. g., the domain and range axioms for the property pz:hasTopping would be added through a pz:AxiomHasTopping template call. Moreover, these axioms can be defined through template calls, e. g., o-owl-ma:DomainRange(pz:hasTopping, pz:Pizza, pz:Topping).

\subsubsection{Template Body Documentation}
To ensure the reuse of templates and the maintenance of their bodies, we propose a template body documentation that is achieved through comments in the OTTR code. This documentation is technical and intended for the ontology engineers (not the domain experts) to explain the functionality and purpose of the statements inside the template bodies.

\subsubsection{Template Library Documentation}
The final documentation step has the goal to create an overview over the template library, including the structure, relationships, and functionality of the templates in the template library to the ontology engineers and domain experts. The documentation encompasses the
following components:
\begin{enumerate}
\def\labelenumi{\alph{enumi})}
\item
  List of all Templates
\item
  Template Inclusion/Call Hierarchy
\item
  Instantiation Order and Naming Guidelines
\item
  Individual Template Documentation
\end{enumerate}

OTTR provides a tool, DocTTR, to automatically generate HTML pages documenting the template library and the individual patterns, covering the list of all templates and the template inclusion/call hierarchy. Here, we use a Semantic MediaWiki (SMW) to organize and store the additional template library documentation along with the templates. We create pages to address the aspects mentioned before, and we make use of the semantic functionalities of the SMW to make the template library and the documentation queryable, e. g., to generate a list of all templates tagged to belong to a certain topic, project, or user. We introduced the naming conventions into the documentation, as we observed that the naming of IRIs and templates held significant importance for the domain experts. It allowed them to easily identify information within the ontology and ensured reusability in other projects. The entire documentation is contained in our dataset published on Zenodo \cite{zenodo}.

\subsubsection{Template Instantiation \& Data Integration}
The ontology is generated from the template library by instantiating templates with data to populate the A-Box, T-Box, or both. Unlike traditional methods that require extensive T-Box detailing before populating the A-Box, our approach builds the T-Box incrementally.

We provide an API for sending template instances to our Semantic MediaWiki (SMW), creating new wiki pages with unique identifiers for each instance, allowing for future data updates. Changes made by ontology engineers to template bodies are easily updated in SMW, with resulting triples automatically being refreshed. Thus, SMW manages templates and instances, takes care of template instantiation, and acts as a triple store. Some template instances include experimental data from our materials science researchers, distinguishing between published and unpublished data, where the latter is kept internal. We use parameters to indicate publication status.

\subsubsection{Details about the Template Library}
\begin{figure}
  \includegraphics[width=\linewidth]{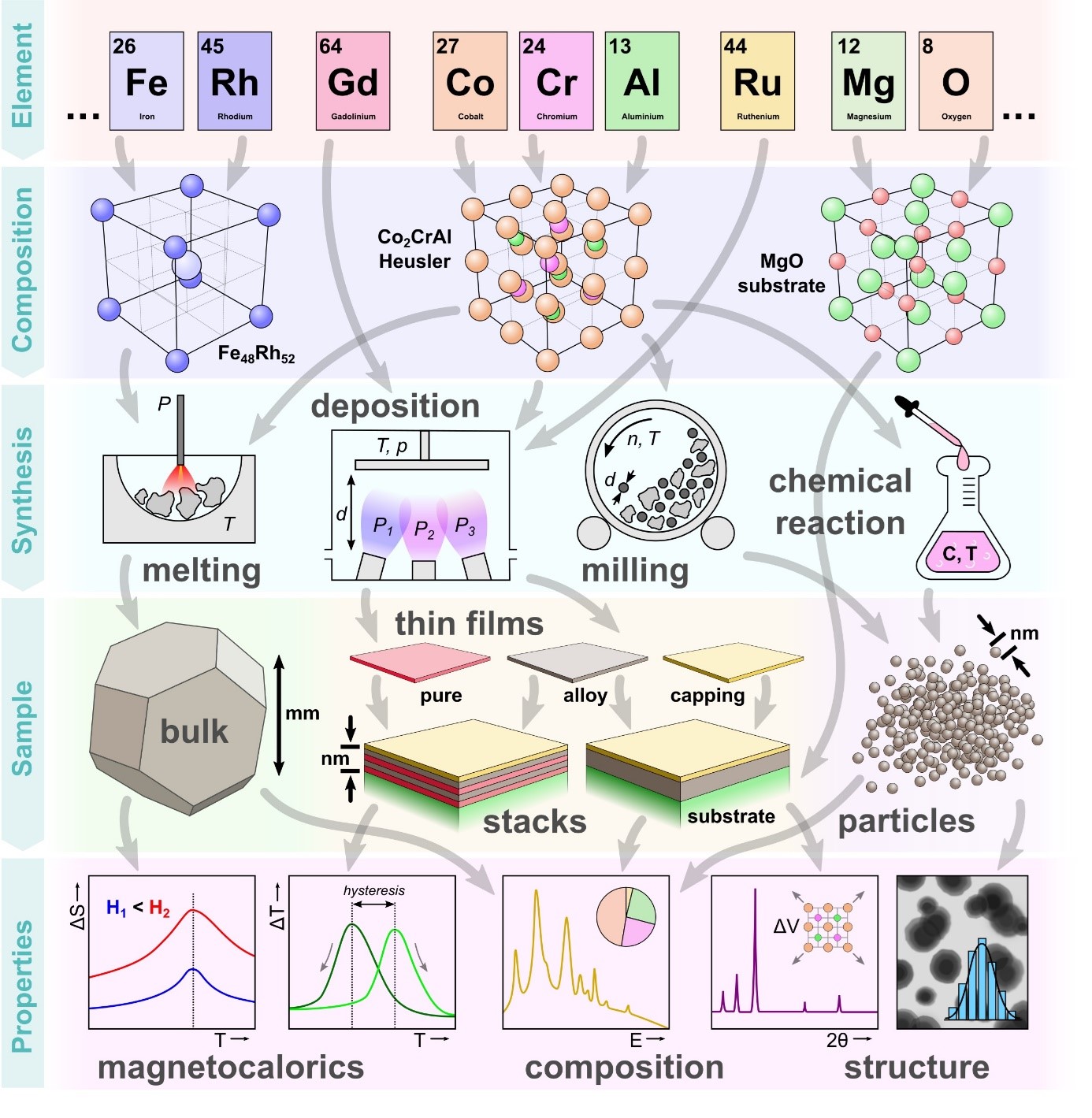}
  \caption{General overview of the DiProMag ontology structure.}
  \label{fig2}
\end{figure}

The following shows the full list of templates that we created for the DiProMag ontology, grouped by the topics composition, synthesis \& treatment, sample, characterization, and goals and ambitions. A general overview about the DiProMag ontology structure is given in \textbf{Figure \ref{fig2}}. 

\uline{Composition}
\begin{itemize}
\item
  dpm:Composition
\item
  dpm:CompositionModification
\item
  dpm:Element
\end{itemize}

\uline{Synthesis \& Treatment}
\begin{itemize}
\item
  dpm:SampleProductionBulkBallMillingJar
\item
  dpm:SampleProductionBulkPress
\item
  dpm:SampleProductionCrushing
\item
  dpm:SampleProductionThinFilmSampleMBEParameters
\item
  dpm:SampleProductionThinFilmSampleMBESource
\item
  dpm:SampleProductionThinFilmSampleSputterParameters
\item
  dpm:SampleProductionThinFilmSampleSputterSource
\item
  dpm:SampleProductionTreatedSample
\item
  dpm:SampleSynthesisBulkArcMelt
\item
  dpm:SampleSynthesisBulkQuartzMelt
\item
  dpm:SampleTreatmentBulkAnneal
\end{itemize}

\uline{Sample}
\begin{itemize}
\item
  dpm:SampleThinFilmSample
\item
  dpm:SampleThinFilmSampleLayer
\item
  dpm:SampleThinFilmSampleSubstrateLayer
\end{itemize}

\begin{itemize}
\item
  dpm:SampleBulk
\end{itemize}

\uline{Characterization}

\begin{itemize}
\item
  dpm:MeasurementDSCTransitionCharacteristics
\item
  dpm:MeasurementEDX
\item
  dpm:MeasurementEDXLine
\item
  dpm:MeasurementEDXMapping
\item
  dpm:MeasurementEDXPoint
\item
  dpm:MeasurementEELS
\item
  dpm:MeasurementEELSCoreLoss
\item
  dpm:MeasurementEELSEnergyLoss
\item
  dpm:MeasurementEELSFittedPeaks
\item
  dpm:MeasurementHeatCapacityDSCZeroField
\item
  dpm:MeasurementMagneticProperties
\item
  dpm:MeasurementMagneticPropertiesEntropyChange
\item
  dpm:MeasurementMagneticPropertiesIsofield
\item
  dpm:MeasurementMagneticPropertiesIsothermal
\item
  dpm:MeasurementMagneticPropertiesTransitionCharacteristics
\item
  dpm:MeasurementXPS
\item
  dpm:MeasurementXPSSubSpectrum
\item
  dpm:MeasurementXPSWideSpectrum
\item
  dpm:MeasurementXRD
\item
  dpm:MeasurementXRF
\item
  dpm:MeasurementXRFResult
\item
  dpm:MeasurementXRR
\item
  dpm:MeasurementXRRMapping
\item
  dpm:Process
\end{itemize}

\uline{Goals \& Ambitions}

\begin{itemize}
\item
  dpm:GoalsAndAmbitions
\end{itemize}

\subsection{Machine Learning on Knowledge Graphs}
As explained above, knowledge graphs can have semantics when terms are used that make entailment rules applicable. This allows to entail facts that are not explicitly contained in the graph. However, these facts are derived via logical deduction, by applying truth-preserving operations
(given that the entailment rules are actually truth-preserving): with the entailment rules from true facts only true facts are derived. Another kind of reasoning is logical induction, with which one can in principle derive facts that could indeed be true, but may also be false. The field of machine learning provides methods that identify and exploit regularities in data and make predictions, for example to predict probable facts. Thus, these methods realize logical induction which can then complement logical deduction carried out via reasoning. We discuss two directions that we explored that exploit regularities in the data.

\subsubsection{Physical Knowledge in Vector Spaces}
How we perceive the world may reflect what the world is like. How we talk (or write) about the world may reflect how we perceive the world. Thus, how we talk about the world may reflect what the world is like. Then, if there is a set of things where we talk about each of these things in a similar way, then these things might in some way be similar. Thus, we can learn about the similarity of things by observing how we talk about these things. This is known as the distributional hypothesis: words that occur in similar contexts tend to have similar meanings \cite{Harris1954}. This would also mean that for a set of things referred to by words that occur in similar contexts, these things are similar / have similar properties. For things relevant to our domain of interest, we could make predictions about physical properties based on whether the words referring to these things occur in similar contexts in a corpus of text. Thus, the question we are interested in is the following: Can we derive physical knowledge from distributional knowledge representations obtained from a corpus of domain-specific texts? Word2vec, by Mikolov and colleagues \cite{mikolov2013}, is a popular family of machine learning approaches that, given a set of words and a corpus of text, via training generate vector representations for words (called embeddings), such that each word is mapped to a vector in a high-dimensional Euclidean space (e. g., $\mathbb{R}^{300}$), such that words that frequently occur in similar contexts are mapped to similar vectors. Here, two vectors are similar if their cosine distance is small. An interesting property of these vector spaces is that they, to some extent, capture the relations between things. For example, it can happen that the vector from the vector of Paris to the vector of France is very similar to the vector from the vector of Berlin to the vector of Germany. Thus, this vector can be interpreted as representing the capital\_of relation. Let's say it is unknown which country Oslo is the capital of, but both Oslo and Norway occur in the vector space, which means that for these two words vectors have been learned. Then, by adding the capital\_of vector to the vector of Oslo, one might find that the vector most similar to the resulting vector is the vector of Norway. Thus, one can obtain the fact from the vector space that Oslo is the capital of Norway. 

t-SNE \cite{vanderMaaten2008} (t-Distributed Stochastic Neighbor Embedding) is a nonlinear dimensionality reduction technique for data visualization. Given a set of embeddings, the approach maps each embedding to a point in 2D or 3D space such that embeddings that are similar (i.e., have low cosine distance) remain close together, while embeddings that are dissimilar (i.e., have high cosine distance) are mapped farther apart. This technique is useful for visually detecting clusters and relationships in the data. 

Tshitoyan and colleagues \cite{Tshitoyan2019} have created a corpus from about 3.3 million scientific abstracts related to materials research and have carried out training using the Word2vec algorithm skip-gram to map words to vectors in $\mathbb{R}^{200}$. They used t-SNE to obtain visualizations which indicate that the vector space contains meaningful clusters, for example that chemically similar elements form clusters.

\begin{figure}
  \includegraphics[width=\linewidth]{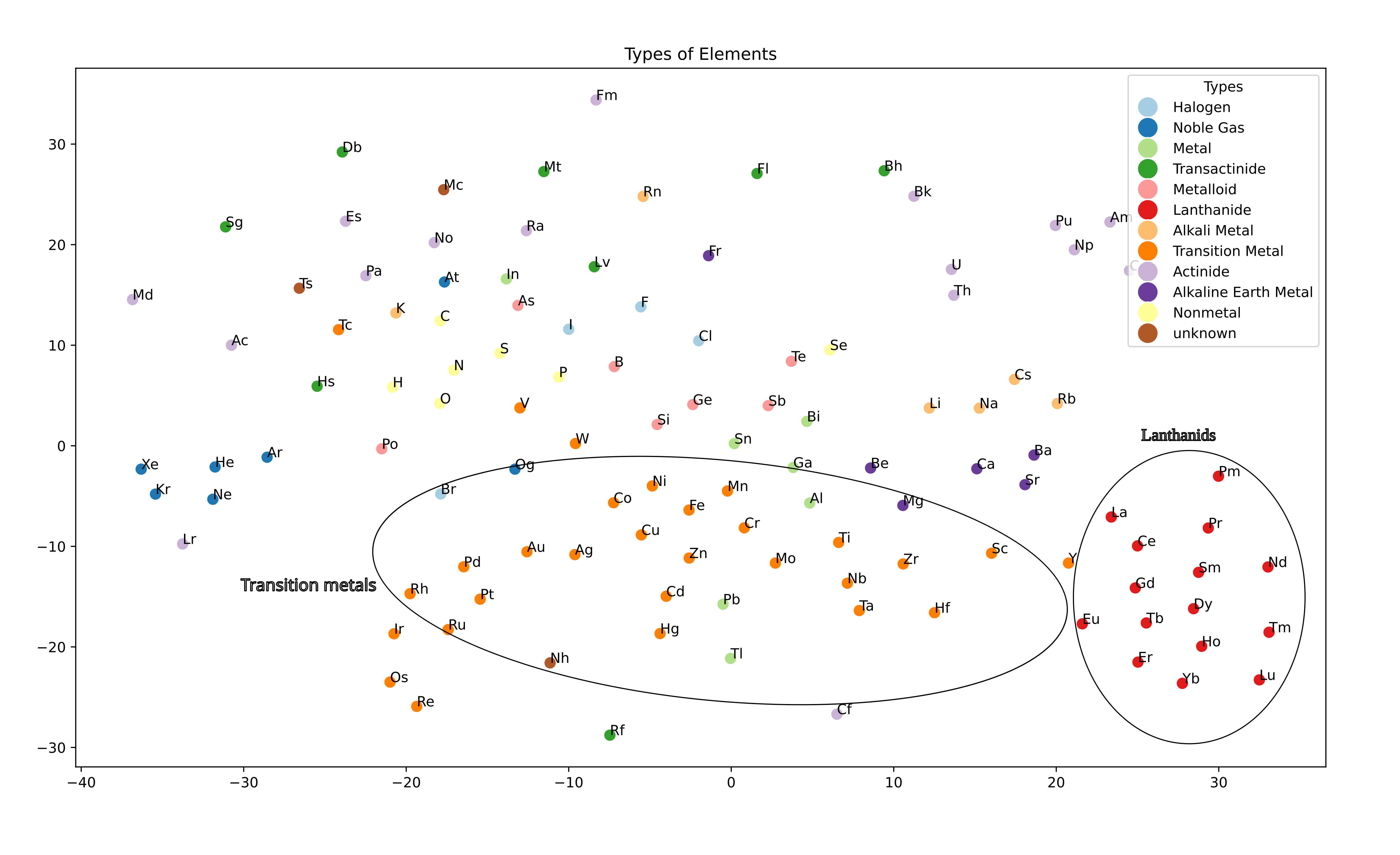}
  \caption{Vector space visualization of chemical elements, colored by element group. One can see that transition metals and lanthanides form clearly visible clusters.}
  \label{fig3}
\end{figure}

We used their vector space to investigate whether physical knowledge relevant to our case is captured in the vector space. This is an interesting question, because if the vector space captures this kind of knowledge, then the vector space can complement our knowledge
graph / we can extend our knowledge graph by deriving facts from the vector space. We used t-SNE to obtain 2D visualizations for sets of chemical elements. We find that chemical elements with similar properties are neighbors in the vector space. In \textbf{Figure \ref{fig3}} one can see that transition metals and lanthanides form clearly visible clusters. Thus, we find that the vector space contains physical knowledge about simple things where that knowledge is relevant to our domain.

\begin{figure}
  \includegraphics[width=\linewidth]{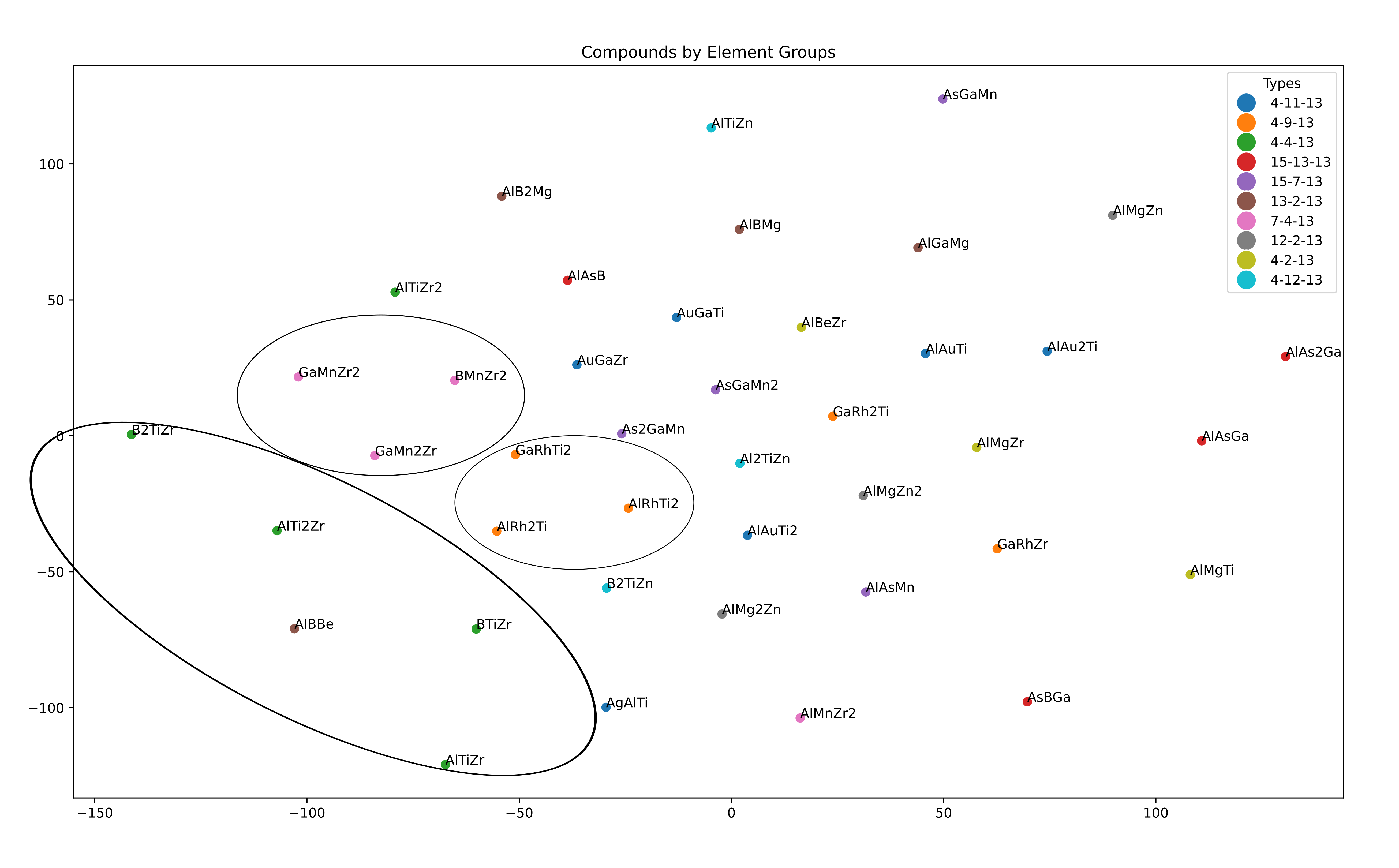} 
  \caption{Vector space visualization of Heusler compounds colored according to in which columns of the periodic table the individual elements that make up the compound are found.}
  \label{fig4}
\end{figure}

An interesting use case where these vector space embeddings and the t-SNE visualizations are used is the following. Consider a set of Heusler compounds where, within some application, a specific Heusler compound is used where some property of the compound is not ideal, e. g., the Martensite start temperature is too high. Then, one would want to observe which compounds occur as neighbors of that compound in the vector space. Ideally, one of the neighbors has the same desired properties with the difference that the Martensite start temperature is lower. Thus, the vector space can help guiding the search for compounds. For example, consider the t-SNE visualization shown in \textbf{Figure \ref{fig4}}. Here, Heusler compounds are shown and colored according to in which columns of the periodic table the individual elements that make up the
compound are found. For example, the combination of columns for the compound AlBMg is 18-2-13. Let's consider the situation that for some application we have carried out experiments with the compounds AlB\textsubscript{2}Mg and AlBMg, which both belong to the category 18-2-13. Then, in the visualization one can see that close-by to the investigated compound there is the compound AlGaMg which might be worth being investigated.

Having words embedded into a vector space enables searching for analogies of the form "A is to B as C is to what?" For example, the solution to the reasoning problem "king is to queen as man is to \_" would be "woman". To solve this task, one can subtract the embedding of man from the embedding of king. The resulting vector could then represent the gender relation. Then, one can subtract that representation from the embedding of queen and find the word which has an embedding that is most similar to the resulting embedding, which should be the word woman. The limitation of this approach is that this vector can represent multiple relations, because often in a corpus of text there are multiple relations described between two entities. However, by creating these tasks, one can probe the vector space for domain-relevant physical knowledge or even discover new knowledge or generate hypotheses that then need to be validated with a closer look into the literature or via experiments. We developed a web application where, given three words A, B, and C, the top-n words are retrieved that are appropriate candidates to fill the blank in the analogical reasoning task "A is to B as C is to \_". 

\subsubsection{Knowledge Graph Completion}
Knowledge graphs are typically incomplete for various reasons such as that they are developed for a particular scope, they are generated from existing incomplete data, the world changes and these changes are not reflected in the knowledge graph, and manual formulation of knowledge is
time-consuming. A family of machine learning approaches aim to exploit regularities in a knowledge graph to predict relations between nodes in the graph. Example approaches are link prediction, node classification, and edge classification.

Link prediction can predict facts not explicitly contained in the graph where these facts cannot be derived via logical deduction. Research has focused mostly on predicting entity-entity relations from a set of entity-entity relations. This is limited because literals which carry important information are ignored. Especially in materials science knowledge graphs important information is stored as literal data, e.g., numerical measurement results. 

Even though some models exist that predict entity-entity relations and incorporate literals for this purpose, it is unclear whether these models are actually better in using numerical literals, or better capable of utilizing the graph structure. We investigate the extent to which link prediction approaches developed for the purpose of utilising literals can indeed make use of them. We created a semi-synthetic dataset and found that many existing models underutilize literal information, even in a setting where the numerical data are
crucial for the prediction. We show that under the established evaluation schema, the performance gains of many models can be attributed to the additional model parameters \cite{blum2024}.

To incorporate literal data, we propose a new approach \cite{blum2023_2} that relies on graph transformations to transform literals into graph structure in such a way that existing Link Prediction methods can leverage the literal information. In particular, we define three
transformations and evaluate them in comparison to state-of-the-art approaches. In most cases, the additional triples generated by our transformations lead to a performance increase.

\section{Implementation of Experimental Results in the Template Structure}
\subsection{Bulk Material}
\begin{figure}
  \includegraphics[width=\linewidth]{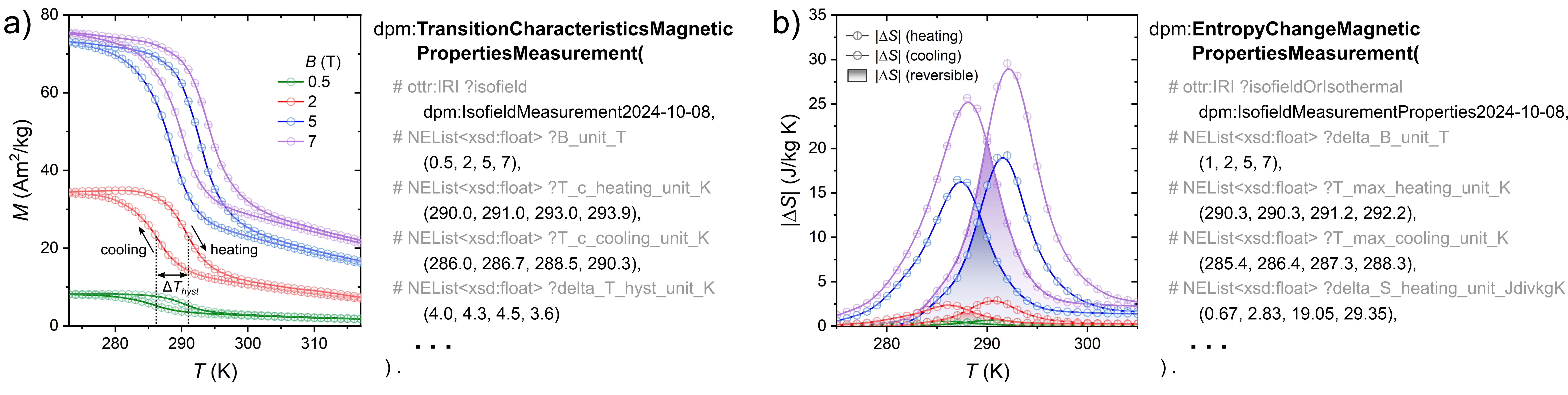}
  \caption{(a) Temperature dependence of the magnetization (M) in the presence of different constant magnetic field (B) during heating and cooling for MnNi\textsubscript{0.8}Co\textsubscript{0.2}Ge\textsubscript{0.97}Al\textsubscript{0.03}. The OTTR template instance captures information about the magnetic phase transition temperatures ($T_C$) for heating and cooling and the thermal hysteresis $\Delta T_{hyst}$ for different magnetic fields $B$. (b) Entropy change ($|\Delta S|$) and associated reversibility ($|\Delta S_{rev}|$) as a function of temperature. The OTTR template instance captures relevant information of $|\Delta S|$, such as maximum entropy changes during heating and cooling, and the associated reversible entropy change.}
  \label{fig56}
\end{figure}

Several bulk magnetic systems have been prepared and characterized for their magnetocaloric properties (MCE). To optimize or improve the MCE, the chemical composition of the stoichiometric compounds/alloys have been modified by substituting new elements in different atomic sites. As
a representative system, the study of MnNi\textsubscript{1-\emph{x}}Co\emph{\textsubscript{x}}Ge\textsubscript{0.97}Al\textsubscript{0.03} (where MnNiGe is the stoichiometric base compound) indicates that the system exhibits a large isothermal entropy change, $\Delta S$ (MCE parameter) \cite{Samanta2023}. The observed large $\Delta S$ is associated with a first-order magnetostructural transition from a low-temperature ferromagnetic orthorhombic to a high-temperature paramagnetic hexagonal phase. The change of crystal structures
(orthorhombic-hexagonal) has been detected in the temperature-dependent X-ray diffraction measurements and related structural parameters (such as, space group, lattice parameters) together with measurement conditions have been successfully incorporated in the developed OTTR templates (not shown here). The values of $\Delta S$ and associated reversibility for the composition MnNi\textsubscript{0.8}Co\textsubscript{0.2}Ge\textsubscript{0.97}Al\textsubscript{0.03} have been estimated from the iso-field magnetization data. The iso-field magnetization data have also been used to determine the transition characteristics (such as, magnetic phase transition temperatures during heating and cooling, thermal hysteresis associated with a first-order phase transition). The plot of iso-field $M(T)$ along with the associated OTTR template instance is shown in \textbf{Figure \ref{fig56} (a)}. The temperature dependence of the entropy change ($|\Delta S|$) and corresponding reversible entropy change ($|\Delta S_{rev}|$) are shown in \textbf{Figure \ref{fig56} (b)} together with the associated OTTR template, in which the relevant information about $|\Delta S|$ (such as, maximal entropy changes during heating and cooling; and associated reversible entropy change) has been structured.

\subsection{Thin films}
\begin{figure}
  \includegraphics[width=\linewidth]{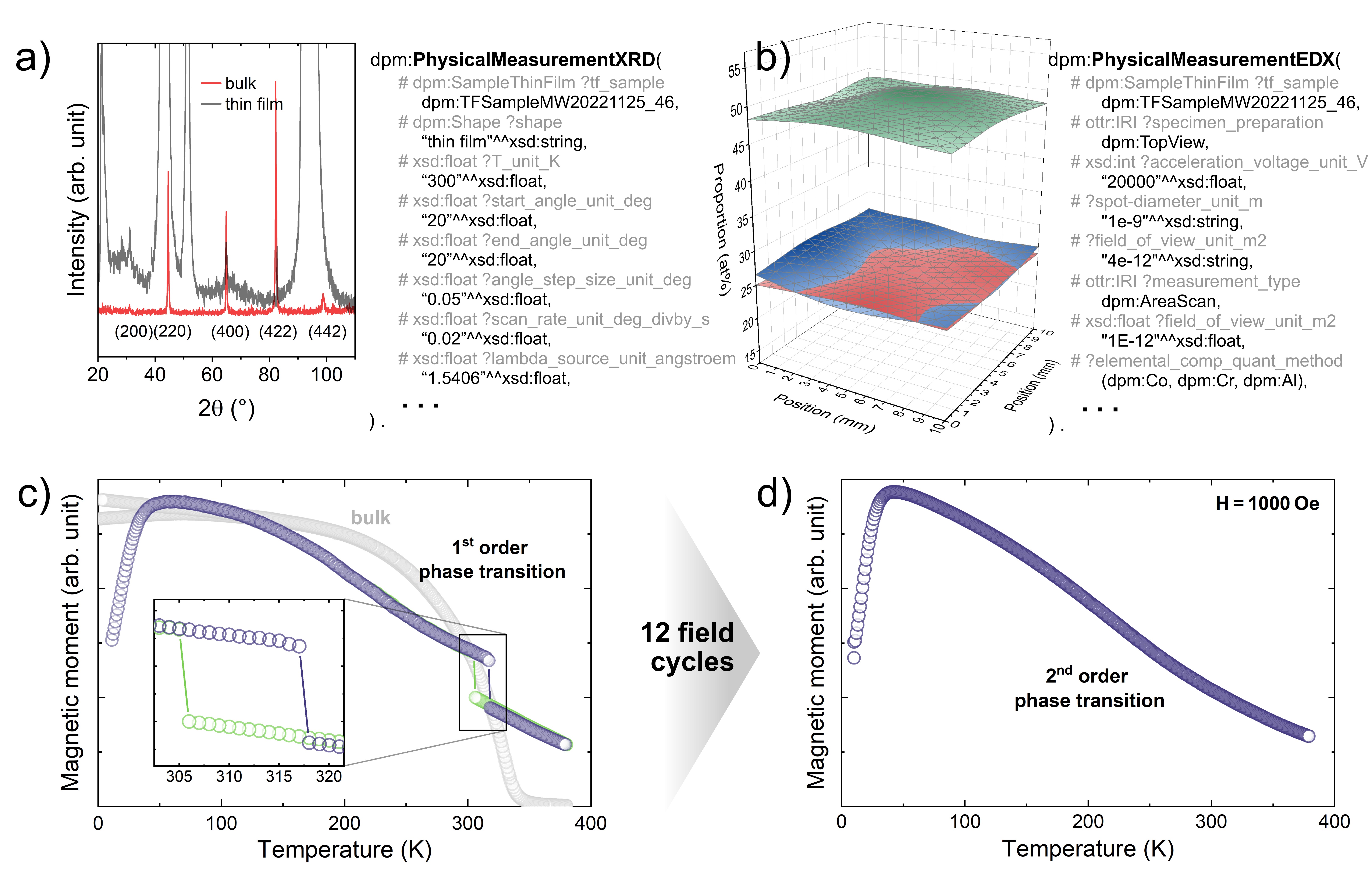}
  \caption{Exemplary results of a thin layer of Co\textsubscript{2}CrAl on a MgO substrate. a) Measurement of the crystal structure by X-ray diffraction (XRD) compared to a bulk sample of the same composition with a section of the corresponding instantiated template. b) Result of many energy dispersive X-ray spectra (EDX) to check the layer homogeneity over the area of the substrate with a section of the corresponding instantiated template. c, d) Measurements of the magnetic moment of a Co\textsubscript{2}CrAl layer as a function of temperature. The additional first-order phase transition at room temperature disappears after 12 magnetic field cycles.}
  \label{fig7}
\end{figure}

Aside from composition and microstructure, geometry is the most important property of a material sample. Thin films are of enormous importance to materials science and solid-state physics, both because they are constitutive for a wide range of microelectronic-based information technologies, and because they enable high-precision and homogeneous material synthesis, providing insights into fundamental physical effects and material properties. In the field of magnetocalorics, thin films also extend beyond basic research to potential applications in chip cooling, microactuators or sensors. Compared to bulk samples, there are a number of similarities in the materials science workflow, but also unique features that should be semantically represented by a holistic ontology. Basic material characteristics such as composition and microstructure, which serve to uniquely identify a material, e. g. an alloy, do not differ significantly from bulk samples. In terms of application and description, most measurement methods demonstrate significant parallels within the context of ontology. This can be efficiently captured through minor extensions to the bulk templates. \textbf{Figure \ref{fig7}a} shows exemplary X-ray diffraction (XRD) measurements of a Co\textsubscript{2}CrAl full Heusler alloy, once for bulk and once for a thin film on a substrate. This example illustrates that the same measurement method follows identical workflows as well as measurement and evaluation methods. The difference, apart from the lower signal intensity, is in the overlay with the substrate (and in some cases additional layers such as capping). While the description of the measurement method is identical to bulk samples, measurement results have to be assigned to the different layers of the thin film stack. Therefore, the layers are treated as individual thin film samples with
their own deposition parameters (currently sputtering and molecular beam epitaxy). Templates of measurement methods refer to templates of material properties, which in turn refer either to templates of single layers or thin film stacks. Both single layers and stacks can refer to
treatment templates such as annealing. Figure \ref{fig7}b shows the lateral homogeneity of the film composition as measured by energy dispersive X-ray spectroscopy (EDX). Again, the procedure is essentially the same as for bulk samples. The template only needs to account for the specific
geometry of the thin film sample. Figure \ref{fig7}c shows the magnetization of the same Co\textsubscript{2}CrAl thin film as a function of temperature. 

The example highlights the geometry-dependence of physical phenomena such as the MCE. Both bulk and thin film show a rather broad 2\textsuperscript{nd} order ferromagnetic to paramagnetic phase
transition. However, in the thin film sample this transition occurs at a lower temperature and over a wider temperature range. In the thin film sample, an additional exceptionally sharp 1\textsuperscript{st} order phase transition associated with a large entropy change was discovered at room temperature. As seen in Figure \ref{fig7}d, this phase transition,
presumably caused by interfacial strain between the film and the substrate, disappeared after 12 field cycles.

\begin{figure}
  \includegraphics[width=\linewidth]{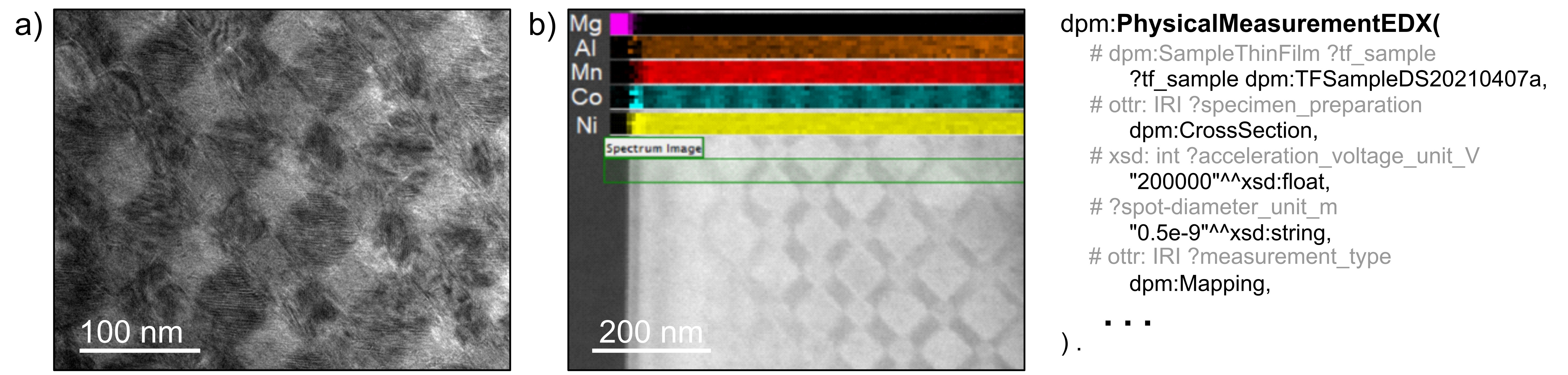}
  \caption{a) TEM image of a cross-sectional lamellae of a NiCoMnAl thin film system showing a 3D nanostructured microstructure. The checkerboard like contrast occurs due to the overlap of different crystal structures, of the austenite phase and 3 martensite variants. b) The EDX mapping is shown with a section of the corresponding instantiated template. The elemental maps show no variation of the composition along with the checkerboard pattern, but the composition variation of the sputtered Heusler thin films can be seen clearly.}
  \label{fig8}
\end{figure}

Strain coupling is also an important reason for the formation of a long-range microstructure in Heusler thin film systems. \textbf{Figure \ref{fig8}a} shows an example of a transmission electron microscopy (TEM) image in which a layered system of non-stoichiometric NiCoMnAl Heusler thin films, crystallized alternately in the martensitic and austenitic phases, exhibits a checkerboard pattern. By forming this 3D microstructure, the thermal hysteresis in this material system could be clearly reduced, making a potential application in magnetocaloric cooling devices more attractive \cite{Ramermann2022, Becker2021}. Since such microstructures are otherwise not captured, the ability to integrate image information into the
ontology is also important. This can be applied together with element-specific analyses such as EDX (see Figure \ref{fig8}b) and electron energy loss spectroscopy (EELS) in order to obtain a comprehensive picture of the thin film sample.

\subsection{Prototyping}
\begin{figure}
  \includegraphics[width=\linewidth]{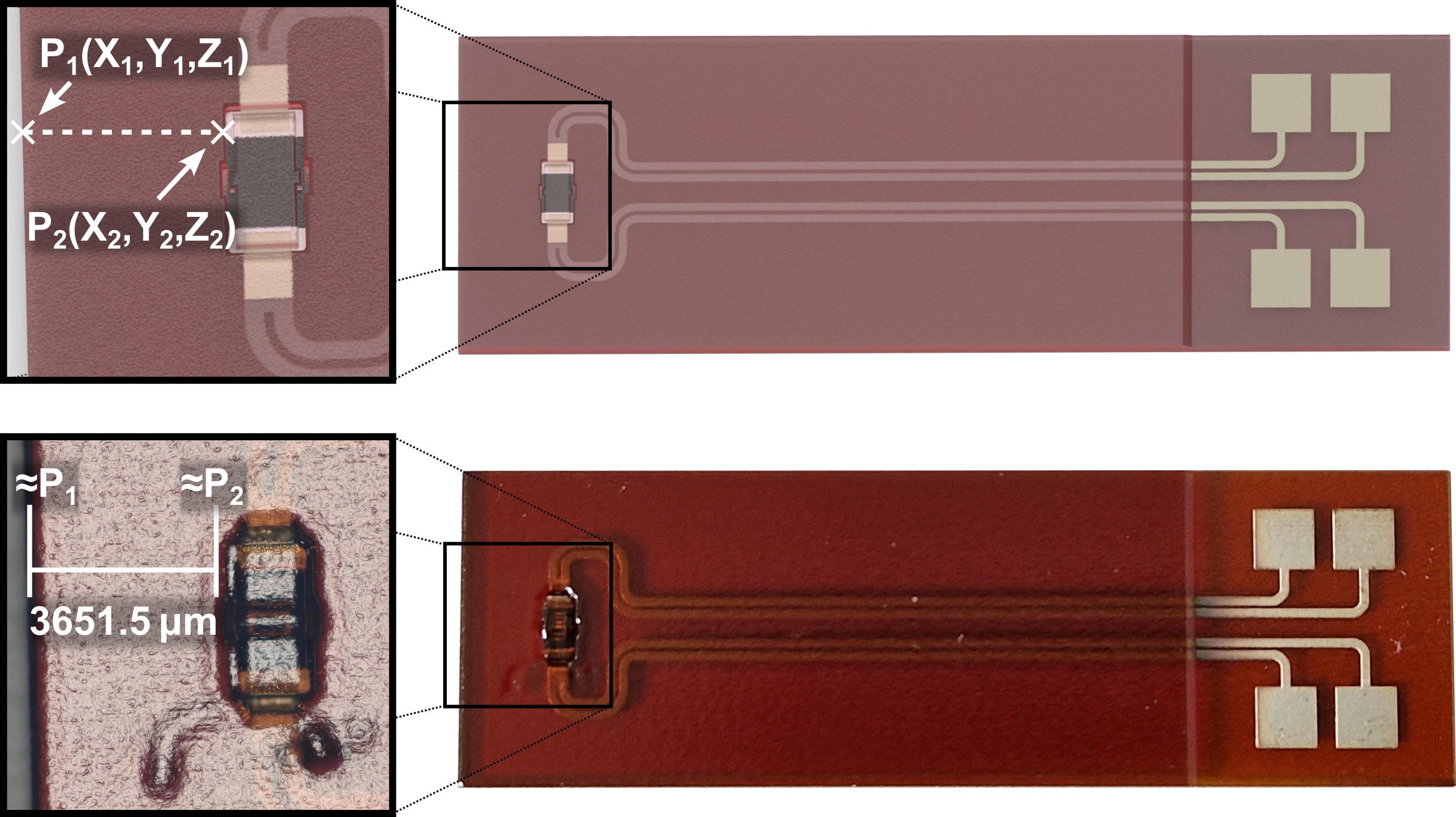}
  \caption{Points with coordinates are defined on the CAD model (top) and these points are relocalized on the prototype (bottom). The measurements are carried out and entered into the template together with the points.}
  \label{fig9}
\end{figure}

In addition to the material properties, the consideration of spatial variables is also important in prototyping. In the application considered here, the prototype is created using computer-aided design (CAD) and is additively manufactured. For example, if a temperature is measured at a certain point or a general check is carried out to determine whether the prototype meets the specifications, the positions must be documented in the ontology. There are various conceivable
possibilities for a deviation between the CAD model and the finished prototype. This will be considered using the example of the position of an imprinted temperature sensor. Traditionally, this would be noted in a sketch in the lab book. However, it should be noted that such a sketch cannot be processed by a machine. To overcome this issue, a machine-processable CAD model with its coordinate system is employed to define positions by coordinates. These positions are relocated on the prototype and then the desired measurement is carried out at one point or between two points. As shown in \textbf{Figure \ref{fig9}}, the length between the two points is measured on the prototype and entered into the template as a measurement between the two points. In this way, for example, it is possible to examine whether the prototype has the desired dimensions. Similarly, positions for temperature measurements are defined or electrical resistances between
two points are measured. 

\begin{figure}
  \includegraphics[width=\linewidth]{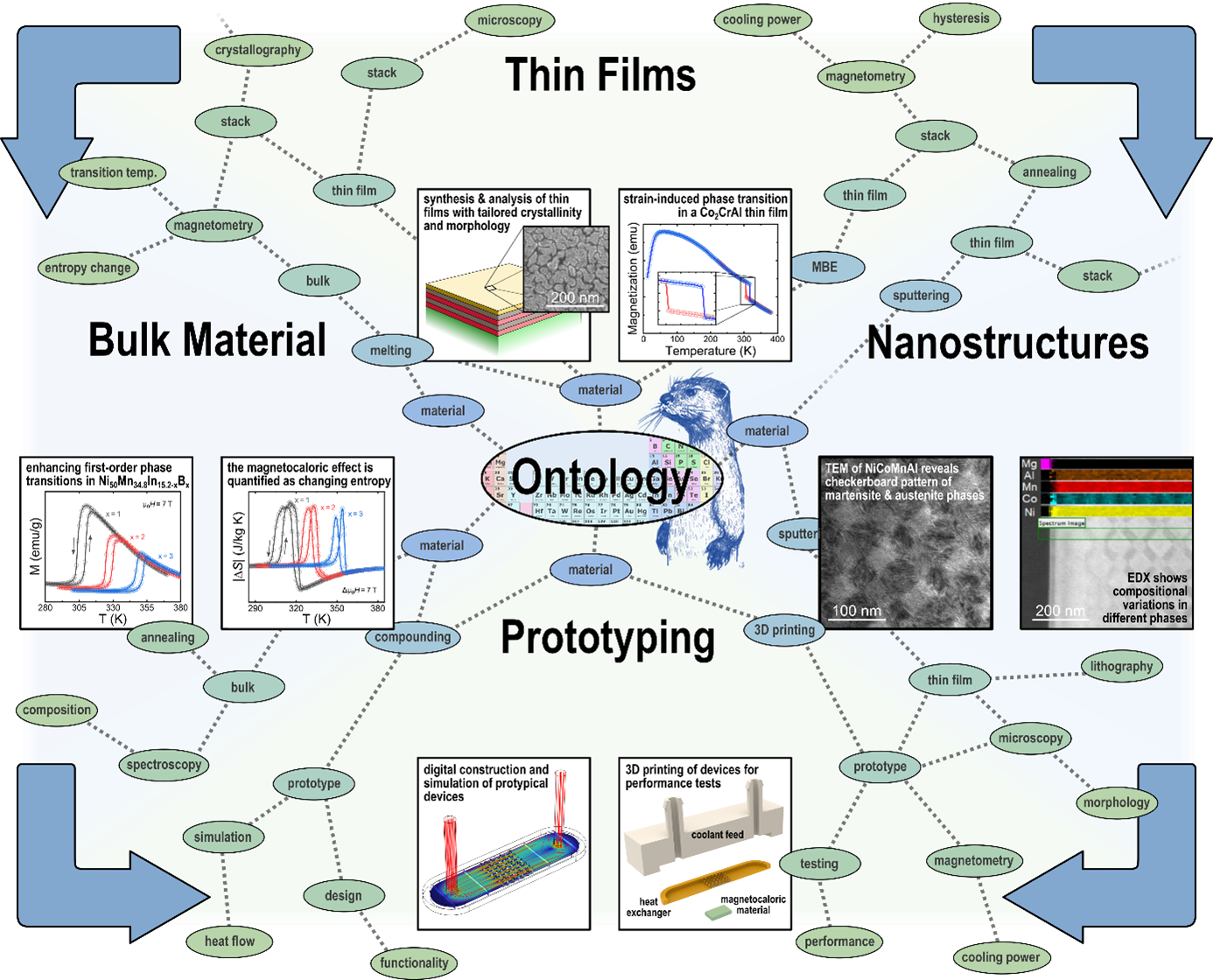}
  \caption{Implementation of experimental results (cf. to \cite{Taake24} for bulk material) into the ontology.}
  \label{fig10}
\end{figure}

An overview of the implementation of experimental results in OTTR-based templates and the  DiProMag ontology is given in \textbf{Figure \ref{fig10}}.

\section{Digital workflows for a combined experimental and theoretical evaluation of the MCE}
Besides the development of OTTR-based templates for the DiProMag ontology, different workflows \cite{Bekemeier2025} have been implemented for the following purposes. Firstly, it encapsulates the data acquisition into ontological structures, as previously outlined. These structures can originate from either experimental or theoretical data sources. Secondly, it facilitates the development of simulation models, making the computational tools more accessible to the user by wrapping them and abstracting away the technical details. This results in an easy-to-use interface. The motivation for creating formalized workflows is to decouple the development of various simulation steps, enabling interdisciplinary teams to collaborate effectively while maintaining a consistent framework. This approach enhances reproducibility by making workflows and their results transferable across groups, promoting collaborative testing and code development.

\begin{figure}
  \includegraphics[width=\linewidth]{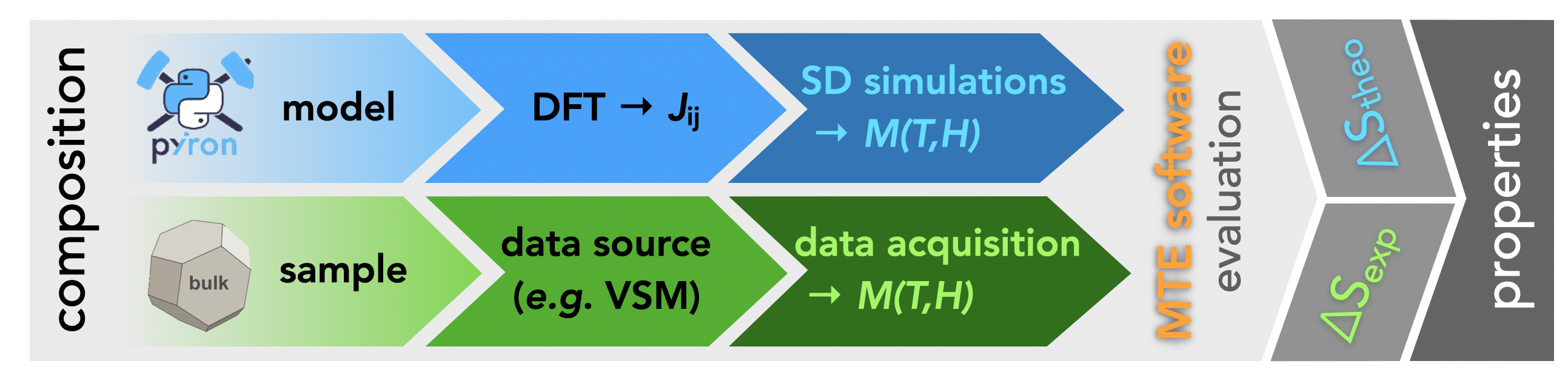}
  \caption{The two workflows, built to combine the simulation and experimental branches. The simulation workflow (top) starts with the atomistic material model, extracts exchange coupling coefficients $J_{ij}$ from DFT results for a spin dynamics simulation, resulting in magnetization curves $M(T,B)$ for the given material. The experimental workflow (below) commences with a sample of the described material, which is characterised by means of measurements. The results of these measurements are automatically canonicalised and recorded in an electronic lab notebook. The result of both workflows is a dataset of magnetization curves that can eventually be used to evaluate the magnetic phase transition and to calculate the magnetic entropy change of the material  (MTE software).}
  \label{fig11}
\end{figure}

We have developed two distinct computational and experimental branches, working towards a unified objective: the evaluation of the magnetocaloric effect of specific materials (e.g. Heusler alloys). Both branches start from the same knowledge point, namely a compound description, and are intended to arrive at the same knowledge point, namely the entropy change of the given magnetocaloric alloy (see \textbf{Figure \ref{fig11}}). 

The computational branch of our workflows consists of \textit{ab initio} calculations for the given material model based on the density functional theory (DFT) and an in-house Markov Chain Monte Carlo (MCMC) spin dynamics (SD) code to simulate a set of magnetization curves. For the experimental branch, we have developed a tool that automates the acquisition and structuring of measurement data from synthesized samples. Both branches ultimately yield data that provide the magnetization curves of the material, whether experimental or simulated, which can then be further analyzed to evaluate the magnetic phase transition. The resulting magnetic entropy change serves as a measure of the magnetocaloric effect in the studied material. The evaluation step can also assess other key parameters, beside the entropy change. Integrating these workflows should ultimately unify experimental characterization and simulation, creating an efficient approach to materials discovery. This will allow seamless sharing of tools and data for both experimental and computational experts, enhancing collaboration across simulation and analysis methods.
\subsection{DFT and extraction of magnetic exchange interaction parameters}
In the theoretical workflow, \textit{ab initio} calculations of the electronic structure using DFT help determine key quantities like total energies, magnetic moments of atoms, and exchange interactions, which are then used in SD simulations to simulate $M(T,B)$, i.e. the magnetization as a function of temperature $T$ and external magnetic field $B$. The workflow is built using the pyiron integrated development environment (IDE) \cite{JANSSEN2019}, taking advantage of the smooth transfer between the integrated tools. By using pyiron, tasks like the creation of atomic structures, which usually means a tedious task depending on the tools used, can be done using built-in libraries and does not have to be done manually. Typically, using such a combination of methods requires in-depth knowledge and considerable experience of working with respective tools, which is usually undertaken by experts. Thanks to the integration of the required tools into the pyiron IDE, this \textit{ab initio} approach becomes also accessible for non-experts and specialists with mostly experimental science background. The simulations are conveniently run in Python Jupyter notebooks. The input of the crystal structures is created using the Atomistic Simulation Environment (ASE), which is integrated into pyiron. For the DFT calculations the VASP software package \cite{VASP} can be used, when a license is available to the user; alternatively, the open-source code SPHInX \cite{BOECK2011543} can be used. Both are pyiron-integrated tools. The calculations can be carried out for various magnetic configurations, whose total energies are required for the calculation of the magnetic exchange parameters. The exchange constants $J_{ij}$ are obtained by regression fitting the total energies for each magnetic configuration to a model Hamiltonian, e.g. the Heisenberg model. Due to the flexibility of Python Jupyter notebooks a customized code part can be easily added, which extracts and fits the total energies automatically. The obtained exchange constants can be further used in the SD simulations, carried out also within the pyiron IDE, all without the need for further manual "translations" between tools, as pyiron data structures can be reused. This workflow can be adjusted for other materials as well. In addition, the use of Python Jupyter notebooks opens broad possibilities for extensions done by each user individually.

\subsection{Monte Carlo spin dynamics simulations}
In order to calculate magnetization curves $M(T,B)$ we have performed classical Monte Carlo spin dynamics simulations \cite{Engelhardt} using our in-house developed code CINOLA \footnote{The availability of the code is possible upon request from the author C.S. Binary versions available in \cite{Schroder_CINOLA_-_Classical_2023})}. The Hamiltonian used for our simulations is given by 
\begin{equation}
\label{Heisenberg}
   H=-\frac12\sum_{ij}J_{ij}\mathbf{m}_i\mathbf{m}_j - \mu_0\mathbf{B}\sum_ig_i\mathbf{m}_i
\end{equation}
Here, the first term describes the Heisenberg exchange interaction between two magnetic moments $\mathbf{m}_i$ and $\mathbf{m}_j$ represented by classical 3D vectors via an exchange-coupling constant $J_{ij}$. The second term defines the interaction of each spin with the external magnetic field $\mathbf{B}$ where $g_i$ is the Landé value which is chosen to be $g_i=2$. Simulations were carried out using three-dimensional cells containing at least 1000 atoms and applying periodic boundary conditions. For each temperature, 106 samples have been used for averaging in order to achieve reasonable accuracy. In our calculations, we consider the interactions between the nearest neighbors and the next-nearest neighbors only. In its original form CINOLA requires the adaption of a detailed configuration file and knowledge of technical details of the code to apply it to a new magnetic system, thus requiring a steep learning curve for new users. However, we developed a user-friendly interface to a subset of CINOLA’s capabilities and wrapped this new CINOLA interface into our pyiron-based workflow \cite{Bekemeier_CINOLA_pyiron_Workflow_2023}. That way, both codes can work in the area of their strengths: The low-level CINOLA code implements the technical details of the necessary algorithms in a computationally optimized way. While the pyiron workflow covers the high-level instructions for setting up the magnetic structure, preprocessing the simulation parameters, calculating exchange parameters, executing the simulation code CINOLA locally or remotely, and postprocessing the results.

\subsection{Automatic data acquisition and canonicalization}
In many laboratories, experimental data and other real-world data still have to be recorded into digital systems manually. This often involves the completion of routine and repetitive tasks that are not the primary objective of the research project. These tasks are instead completed to enable the achievement of the final goal. It is not uncommon for researchers working in the same workspace to employ different methodologies for handling their data, despite the fact that the data in question have been collected in the same environment, utilising the same equipment, and will subsequently be subjected to the same post-processing procedures by the same individuals. This results in a multitude of unique instances where the acquired data must be transferred and transformed when researchers collaborate on their data or disseminate their tools.

Furthermore, we have developed a solution on how to automate the tasks of data acquisition and by doing so reduce human intervention in mundane tasks such as data handling, homogenize the way data is handled and therefore allow seamless ingestion and transformation of experimental data for use in simulations and further analysis.
\begin{figure}
  \includegraphics[width=\linewidth]{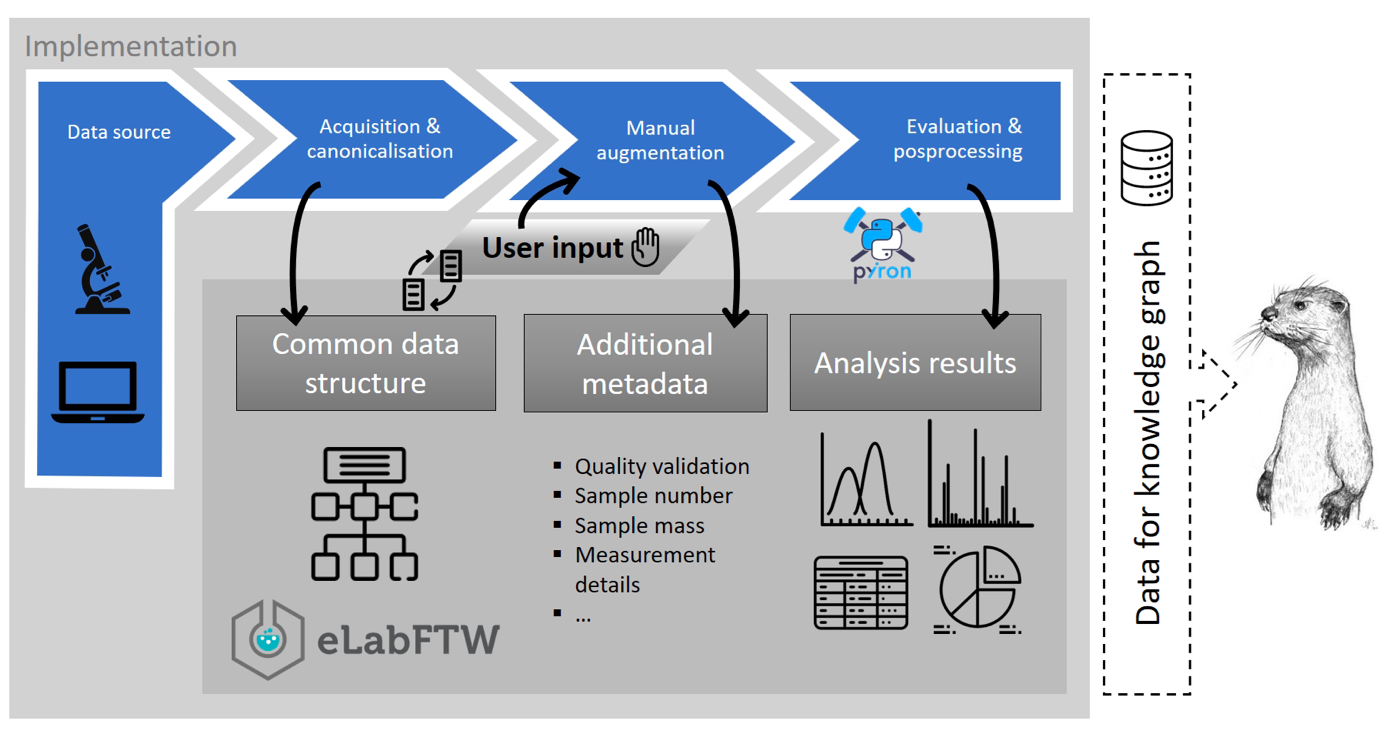}
  \caption{Acquisition of experimental data entails the collection and standardization of magnetization curves from a proprietary format (obtained from a vibrating sample magnetometer (VSM), Quantum Design MPMS) and their incorporation into a centralized electronic laboratory notebook. The data is automatically transformed into a structured format based on the ontology presented in this work based on OTTR-templates and stored for convenient access.}
  \label{fig12}
\end{figure}

In the conventional laboratory procedure examined in this study, the data is collected on a  computer that is linked to the measuring apparatus. The files are then manually transferred to other devices or storage locations for further processing. Once a measurement is complete, the lab user manually copies the resulting file to a designated destination on a shared network. The automated acquisition process is then triggered by the newly identified file in that directory. The raw measurement file remains in a format specific to the vendor or measurement device, with terminology and definitions unique to that vendor or device. In the case of the Quantum Design MPMS VSM magnetometer, this constitutes a vendor-specific alteration of the CSV file type convention. Triggered by the creation of the new file in the shared file directory, this raw measurement file is read by the Python-based Data Acquisition Pipeline and transferred into an eLabFTW experiment entry in modularized steps.

A selection process was conducted involving laboratory personnel and pipeline developers. Through this process, certain standards were agreed upon that align with typical laboratory procedures while also facilitating structured, automated processes. Consequently, the file path can be employed to ascertain the appropriate experiment title, which was previously predominantly encoded into the filename by laboratory users through the use of individualized patterns. This practice has the potential to hinder automation.
The metadata associated with the measurement process, which are included in the raw file, are converted into a canonical form that has been previously agreed upon within the context of the project. This agreement is comprised of two distinct components. Firstly, the semantics of the metadata were identified and codified during the ontology development process, as described above. Secondly, throughout the aforementioned selection process, standard syntax and units were defined, thereby establishing the manner in which manual inputs to the measurement apparatus must be provided by laboratory users. These inputs are then directly transferred to the resulting raw measurement file by the device. This process eliminates the file-type specific details, terminology, and syntax unique to the measurement device, as well as the lab user-specific input field units. The metadata are then converted into a structure and naming convention that aligns with the previously developed ontology.

In addition to the data that is automatically ingested by the process described above, there also exists data that cannot be automatically acquired from the measurement process or its environment. This is typically data that is stored in the laboratory user's memory or in their notes. An example of this would be the orientation of the measured sample when it was placed into the measurement device. It is necessary to import this data into the system as a separate entity. In order to achieve this, the electronic lab notebook eLabFTW was utilized as an accessible user interface for laboratory personnel, who were already somewhat familiar with it due to its resemblance to the conventional paper-based lab notebook. In this regard, the selection process enabled the identification of the metadata that will exist outside the measurement device's output file, as well as the manner in which it should be integrated into the automated process. The agreed-on data fields were subsequently encoded into eLabFTW's "extra\_field" functionality, which enables the specification of supplementary metadata fields that will be displayed in a structured manner alongside the primary text of an experiment entry. 
The additional fields, together with the data that has been automatically acquired and transformed previously, are then uploaded to an eLabFTW instance and extended with the additional metadata of the laboratory personnel.
To guarantee the complete reproducibility of all measured data, it is essential to upload not only the processed data to eLabFTW, but also the raw output file from the measurement device and the extracted, unprocessed header fields from that file. The developed process is thus far limited to the automatic acquisition of metadata. The actual measurement data is transformed into a pure CSV format and uploaded to the eLabFTW experiment entry without further processing. This facilitates the import of data into tools for subsequent processing and analysis, avoiding the necessity for translating the vendor-specific file format into a standard CSV format that can be utilized by other software. Nevertheless, the subsequent processing of the data can and should be tailored to align with the specific requirements of the researchers' or laboratory's subsequent analysis. This is, however, beyond the scope of the present work.

\subsection{Evaluation of magnetic phase transition}
The entropy change $\Delta S$ can be calculated using \cite{Planes}:
\begin{equation}
\label{calc}
   \Delta S_M = \int_{B_i}^{B_f} \left( \frac{\partial M}{\partial T} \right)_B dB
\end{equation}
where $B_i$ and $B_f$ are the initial and final magnetic fields, respectively, $M$ is the magnetization, and $T$ is the temperature. Equation \ref{calc} expresses how the magnetization changes with temperature under an applied magnetic field, i.e. reflecting the material's ability to absorb or release heat through the change in magnetic order. As illustrated by this equation, the data utilized for evaluation should exhibit consistent field values (which is the case for isofield measurements). Furthermore, common temperature values are necessary for the integration to be performed. In the context of simulated data, these properties are typically already defined by the parameters utilized in the simulation. In contrast, if one is dealing with experimental data, this is not the default case; however, it can be achieved without difficulty through the careful interpolation of the data.
\begin{figure}
  \includegraphics[width=\linewidth]{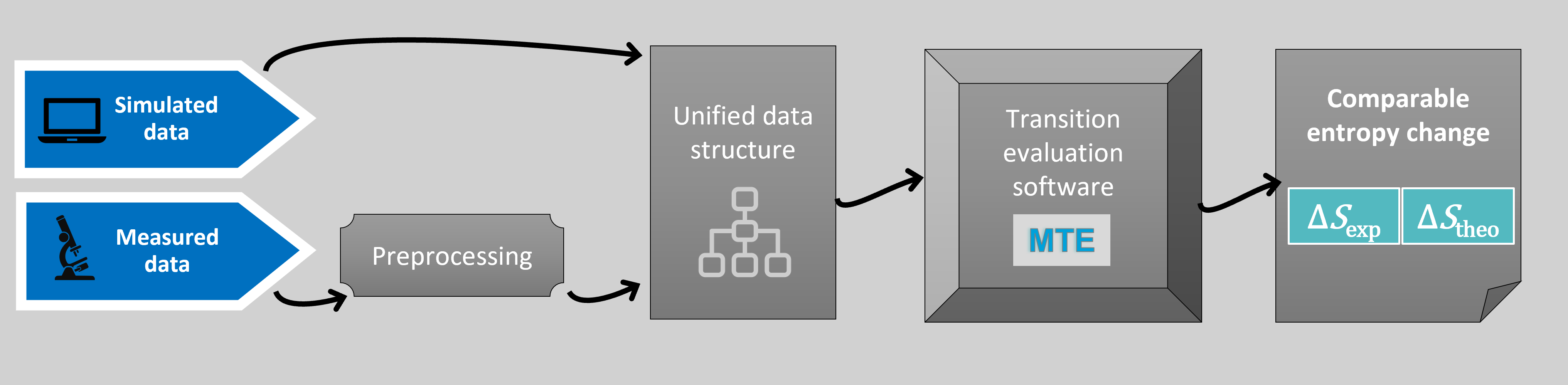}
  \caption{Data and software streams for the entropy change calculation.}
  \label{fig13}
\end{figure}
From a software tool perspective, this presents a significant advantage, as both simulated data and interpolated experimental data are derived from the same fundamental basis for subsequent calculations. This allows for the treatment of these data sets in an identical manner (see \textbf{Figure \ref{fig13}}).

\subsection{Automatic execution of workflows based on central data store}
To further streamline the use of software tools like the one described in the previous sections, a prototype solution was developed. This solution aims to hide the technical details of the computational environment, the execution of the code, and the data transfer between the various components of the software system. The prototype assumes that the workflow to be executed should be defined as a Python Jupyter notebook. This Jupyter notebook does not require a specific format for execution, as we anticipate that the majority of such workflows will be developed by researchers for their own purposes. The integration of other researchers' workflows into the automated system should not be constrained by excessive requirements regarding workflow format. Accordingly, the form of the workflow notebook is subject to minimal requirements. Notebook cells, which encompass inputs to the workflow or outputs of the workflow, must be marked in a particular manner. It is further assumed that the necessary input data can be obtained from an eLabFTW experiment entry, and that the workflow outputs should be uploaded back into the same entry. In the typical configuration, the workflow notebooks are stored on a central server, ensuring that all laboratory users have access to these workflows. To execute the desired workflow, the laboratory user simply needs to select the appropriate workflow, choose the relevant eLabFTW experiment, and map the available data from the experiment entry to the input cells, which were previously marked in the workflow notebook. The necessary data are automatically loaded from the eLabFTW instance and made available to the workflow execution environment. The automated system also supervises the workflow environment and execution, executing the work-flow and extracting the previously marked outputs. The aforementioned outputs are then automatically uploaded to eLabFTW and appended to the original experiment entry. This prototype does not include all the necessary features to make the process fully transparent to laboratory users. However, it can be used to demonstrate the potential benefits of automation and data structuring to lab users, and to provide context for the initial work required to achieve such automation.

In conclusion, the integration of these workflows and automation tools will facilitate the convergence of the experimental and simulation processes, ultimately enabling a seamless integration of both aspects of the materials discovery process, namely experiments and simulation. This integration will not only enhance the presentation of findings in academic publications but also streamline the discovery process itself, enabling experimental and computational experts to leverage each other's tools and data, and to readily combine diverse simulation and analysis tools. This can only be achieved by ensuring that the data is findable, accessible, interoperable, and reusable (FAIR), and that the workflows used to process it are also FAIR.

\section{Conclusion}
We developed an OTTR-centric ontology engineering methodology, developed tool support for template prototyping and validation in the form of an extension to Semantic MediaWiki, and applied the methodology to develop the DiProMag ontology and to capture research results. We made the tool and the ontology available to the community. We believe that the ontology and the templates can be useful to other projects, without or with minor modifications. Furthermore, we found that embedding spaces can contain interesting physical knowledge and we developed machine
learning approaches to predict facts in knowledge graphs, i. e., knowledge graph completion methods.

In the future, when more data samples are available in the ontology, knowledge graph completion methods can better identify regularities, thus improve the predictions. The more these ontologies are used and research data and results are made available in semantic form, the more material sciences research can benefit, thus fostering innovation.

We have used OTTR-based templates and workflows to unify experimental and computational methods for evaluating the magnetocaloric effect in materials. These workflows streamline data acquisition and simulation by encapsulating experimental and theoretical information in ontological structures and offering user-friendly interfaces to computational tools. Ultimately, the implementation of these workflows supports unified material characterization, seamless sharing of tools and data, enhanced collaboration across disciplines, FAIR principles, and advances materials discovery. A prototype solution integrating Python Jupyter notebooks with eLabFTW was developed to simplify the use of computational tools by hiding technical details and automating data transfer between software components. 

In the evolving landscape of materials science, integrating experimental and computational approaches is crucial for advancing discoveries. The DiProMag ontology serves as a prime example of how formalized workflows can bridge these domains, enabling seamless collaboration and efficient analysis. By encapsulating data into ontological structures and simplifying the user interface for computational tools, we can empower interdisciplinary teams to explore the magnetocaloric properties of materials like Heusler alloys with unprecedented efficiency.

\medskip
\textbf{Acknowledgements} \par 
This research has been conducted within the DiProMag (Digitization of a process chain for the synthesis, characterization and prototypical application of magnetocaloric alloys) project as part of the "Initiative MaterialDigital", targeting digital transformation in Materials research. The authors thank the German Federal Ministry of Research, Technology and Space (BMFTR) for financial support through project funding grants no. 3XP5094E (Federal Institute for Materials Research and Testing), 13XP5120A (Bielefeld University of Applied Sciences and Arts), 13XP5120B (Bielefeld University).


%
\bibliographystyle{IEEEtran}
\bibliography{bibliography}

@article{Giauque1933,
  title = {Attainment of Temperatures Below 1\ifmmode^\circ\else\textdegree\fi{} Absolute by Demagnetization of ${\mathrm{Gd}}_{2}$${(\mathrm{S}{\mathrm{O}}_{4})}_{3}$\ifmmode\cdot\else\textperiodcentered\fi{}8${\mathrm{H}}_{2}$O},
  author = {Giauque, W. F. and MacDougall, D. P.},
  journal = {Phys. Rev.},
  volume = {43},
  issue = {9},
  pages = {768--768},
  numpages = {0},
  year = {1933},
  month = {May},
  publisher = {American Physical Society},
  doi = {10.1103/PhysRev.43.768},
  url = {https://link.aps.org/doi/10.1103/PhysRev.43.768}
}

@article{Debye1926,
author = {Debye, P.},
title = {Einige Bemerkungen zur Magnetisierung bei tiefer Temperatur},
journal = {Annalen der Physik},
volume = {386},
number = {25},
pages = {1154-1160},
doi = {https://doi.org/10.1002/andp.19263862517},
url = {https://onlinelibrary.wiley.com/doi/abs/10.1002/andp.19263862517},
eprint = {https://onlinelibrary.wiley.com/doi/pdf/10.1002/andp.19263862517},
year = {1926}
}

@article{Brown1976,
    author = {Brown, G. V.},
    title = {Magnetic heat pumping near room temperature},
    journal = {Journal of Applied Physics},
    volume = {47},
    number = {8},
    pages = {3673-3680},
    year = {1976},
    month = {08},
    issn = {0021-8979},
    doi = {10.1063/1.323176},
    url = {https://doi.org/10.1063/1.323176},
    eprint = {https://pubs.aip.org/aip/jap/article-pdf/47/8/3673/18373880/3673\_1\_online.pdf},
}

@article{Studer1998,
title = {Knowledge engineering: Principles and methods},
journal = {Data \& Knowledge Engineering},
volume = {25},
number = {1},
pages = {161-197},
year = {1998},
issn = {0169-023X},
doi = {https://doi.org/10.1016/S0169-023X(97)00056-6},
url = {https://www.sciencedirect.com/science/article/pii/S0169023X97000566},
author = {Rudi Studer and V.Richard Benjamins and Dieter Fensel}
}

@article{Hogan2021,
author = {Hogan, Aidan and Blomqvist, Eva and Cochez, Michael and D’amato, Claudia and Melo, Gerard De and Gutierrez, Claudio and Kirrane, Sabrina and Gayo, Jos\'{e} Emilio Labra and Navigli, Roberto and Neumaier, Sebastian and Ngomo, Axel-Cyrille Ngonga and Polleres, Axel and Rashid, Sabbir M. and Rula, Anisa and Schmelzeisen, Lukas and Sequeda, Juan and Staab, Steffen and Zimmermann, Antoine},
title = {Knowledge Graphs},
year = {2021},
issue_date = {May 2022},
publisher = {Association for Computing Machinery},
address = {New York, NY, USA},
volume = {54},
number = {4},
issn = {0360-0300},
url = {https://doi.org/10.1145/3447772},
doi = {10.1145/3447772},
journal = {ACM Comput. Surv.},
month = jul,
articleno = {71},
numpages = {37}
}

@article{Uschold1996,
author = {Uschold, Michael and Gr\"uninger, Michael},
year = {1996},
month = {01},
pages = {},
title = {Ontologies: Principles, methods and applications},
volume = {11},
journal = {The Knowledge Engineering Review}
}

@inbook{Abecker2009,
author = {Abecker, Andreas and van Elst, Ludger},
bookTitle="Handbook on ontologies",
year = {2009},
month = {05},
pages = {713-734},
title = {Ontologies for Knowledge Management},
publisher="Springer Berlin Heidelberg",
address="Berlin, Heidelberg",
isbn = {978-3-540-70999-2},
doi = {10.1007/978-3-540-92673-3_32}
}

@Inbook{Horrocks2013,
  title={What are ontologies good for?},
  author={Horrocks, Ian},
  booktitle={Evolution of semantic systems},
  pages={175--188},
  year={2013},
  publisher={Springer}
}

@Article{skjaeveland2024,
  author =	{Skj{\ae}veland, Martin Georg and Karlsen, Leif Harald},
  title =	{{The Reasonable Ontology Templates Framework}},
  journal =	{Transactions on Graph Data and Knowledge},
  pages =	{5:1--5:54},
  ISSN =	{2942-7517},
  year =	{2024},
  volume =	{2},
  number =	{2},
  publisher =	{Schloss Dagstuhl -- Leibniz-Zentrum f{\"u}r Informatik},
  address =	{Dagstuhl, Germany},
  URL =		{https://drops.dagstuhl.de/entities/document/10.4230/TGDK.2.2.5},
  URN =		{urn:nbn:de:0030-drops-225896},
  doi =		{10.4230/TGDK.2.2.5}
}

@InProceedings{skjaeveland2018,
author="Skj{\ae}veland, Martin G.
and Lupp, Daniel P.
and Karlsen, Leif Harald
and Forssell, Henrik",
editor="Vrande{\v{c}}i{\'{c}}, Denny
and Bontcheva, Kalina
and Su{\'a}rez-Figueroa, Mari Carmen
and Presutti, Valentina
and Celino, Irene
and Sabou, Marta
and Kaffee, Lucie-Aim{\'e}e
and Simperl, Elena",
title="Practical Ontology Pattern Instantiation, Discovery, and Maintenance with Reasonable Ontology Templates",
booktitle="The Semantic Web -- ISWC 2018",
year="2018",
publisher="Springer International Publishing",
address="Cham",
pages="477--494",
isbn="978-3-030-00671-6"
}

@misc{blum2023,
      title={Insights from an OTTR-centric Ontology Engineering Methodology}, 
      author={Moritz Blum and Basil Ell and Philipp Cimiano},
      year={2023},
      eprint={2309.13130},
      archivePrefix={arXiv},
      primaryClass={cs.DB},
      url={https://arxiv.org/abs/2309.13130}, 
}

@article{Bayerlein2025,
author = {Bayerlein, Bernd and Waitelonis, J\"org and Birkholz, Henk and Jung, Matthias and Schilling, Markus and v. Hartrott, Philipp and Bruns, Marian and Schaarschmidt, J\"org and Beilke, Kristian and Mutz, Marcel and Nebel, Vincent and K\"oniger, Veit and Beran, Lisa and Kraus, Tobias and Vyas, Akhilesh and Vogt, Lars and Blum, Moritz and Ell, Basil and Chen, Ya-Fan and Waurischk, Tina and Thomas, Akhil and Durmaz, Ali Riza and Ben Hassine, Sahar and Fresemann, Carina and Dziwis, Gordian and Beygi Nasrabadi, Hossein and Hanke, Thomas and Telong, Melissa and Pirskawetz, Stephan and Kamal, Mohamed and Bjarsch, Thomas and P\"ahler, Ursula and Hofmann, Peter and Leemhuis, Mena and \"ozçep, \"ozg\"ur L. and Meyer, Lars-Peter and Skrotzki, Birgit and Neugebauer, J\"org and Wenzel, Wolfgang and Sack, Harald and Eberl, Chris and Portella, Pedro Dolabella and Hickel, Tilmann and M\"adler, Lutz and Gumbsch, Peter},
title = {Concepts for a Semantically Accessible Materials Data Space: Overview over Specific Implementations in Materials Science},
journal = {Advanced Engineering Materials},
volume = {27},
number = {8},
pages = {2401092},
doi = {https://doi.org/10.1002/adem.202401092},
url = {https://advanced.onlinelibrary.wiley.com/doi/abs/10.1002/adem.202401092},
eprint = {https://advanced.onlinelibrary.wiley.com/doi/pdf/10.1002/adem.202401092},
year = {2025}
}

@article{Harris1954,
author = {Zellig S. Harris},
title = {Distributional Structure},
journal = {WORD},
volume = {10},
number = {2-3},
pages = {146--162},
year = {1954},
publisher = {Routledge},
doi = {10.1080/00437956.1954.11659520},


URL = { 
    
        https://doi.org/10.1080/00437956.1954.11659520
    
    

},
eprint = { 
    
        https://doi.org/10.1080/00437956.1954.11659520
    
    

}

}

@misc{mikolov2013,
      title={Efficient Estimation of Word Representations in Vector Space}, 
      author={Tomas Mikolov and Kai Chen and Greg Corrado and Jeffrey Dean},
      year={2013},
      eprint={1301.3781},
      archivePrefix={arXiv},
      primaryClass={cs.CL},
      url={https://arxiv.org/abs/1301.3781}, 
}

@article{vanderMaaten2008,
  author  = {Laurens van der Maaten and Geoffrey Hinton},
  title   = {Visualizing Data using t-SNE},
  journal = {Journal of Machine Learning Research},
  year    = {2008},
  volume  = {9},
  number  = {86},
  pages   = {2579--2605},
  url     = {http://jmlr.org/papers/v9/vandermaaten08a.html}
}

@Article{Tshitoyan2019,
author={Tshitoyan, Vahe
and Dagdelen, John
and Weston, Leigh
and Dunn, Alexander
and Rong, Ziqin
and Kononova, Olga
and Persson, Kristin A.
and Ceder, Gerbrand
and Jain, Anubhav},
title={Unsupervised word embeddings capture latent knowledge from materials science literature},
journal={Nature},
year={2019},
month={Jul},
day={01},
volume={571},
number={7763},
pages={95-98},
issn={1476-4687},
doi={10.1038/s41586-019-1335-8},
url={https://doi.org/10.1038/s41586-019-1335-8}
}

@misc{blum2024,
      title={Numerical Literals in Link Prediction: A Critical Examination of Models and Datasets}, 
      author={Moritz Blum and Basil Ell and Hannes Ill and Philipp Cimiano},
      year={2024},
      eprint={2407.18241},
      archivePrefix={arXiv},
      primaryClass={cs.LG},
      url={https://arxiv.org/abs/2407.18241}, 
}

@inproceedings{blum2023_2,
author = {Blum, Moritz and Ell, Basil and Cimiano, Philipp},
title = {Exploring the impact of literal transformations within Knowledge Graphs for Link Prediction},
year = {2023},
isbn = {9781450399876},
publisher = {Association for Computing Machinery},
address = {New York, NY, USA},
url = {https://doi.org/10.1145/3579051.3579069},
doi = {10.1145/3579051.3579069},
booktitle = {Proceedings of the 11th International Joint Conference on Knowledge Graphs},
pages = {48–54},
numpages = {7},
location = {Hangzhou, China},
series = {IJCKG '22}
}

@article{Samanta2023,
doi = {10.1088/2515-7655/acf957},
url = {https://dx.doi.org/10.1088/2515-7655/acf957},
year = {2023},
month = {sep},
publisher = {IOP Publishing},
volume = {5},
number = {4},
pages = {044002},
author = {Samanta, Tapas and Taake, Chris and Bondzio, Laila and Caron, Luana},
title = {Entropy change reversibility in MnNi1−xCoxGe0.97Al0.03 near the triple point},
journal = {Journal of Physics: Energy}
}

@Article{Ramermann2022,
AUTHOR = {Ramermann, Daniela and Becker, Andreas and B\"uker, Bj\"orn and H\"utten, Andreas and Ennen, Inga},
TITLE = {Nano Scaled Checkerboards: A Long Range Ordering in NiCoMnAl Magnetic Shape Memory Alloy Thin Films with Martensitic Intercalations},
JOURNAL = {Applied Sciences},
VOLUME = {12},
YEAR = {2022},
NUMBER = {3},
ARTICLE-NUMBER = {1748},
URL = {https://www.mdpi.com/2076-3417/12/3/1748},
ISSN = {2076-3417},
DOI = {10.3390/app12031748}
}

@Article{Becker2021,
AUTHOR = {Becker, Andreas and Ramermann, Daniela and Ennen, Inga and B\"uker, Bj\"orn and Matalla-Wagner, Tristan and Gottschalk, Martin and H\"utten, Andreas},
TITLE = {The Influence of Martensitic Intercalations in Magnetic Shape Memory NiCoMnAl Multilayered Films},
JOURNAL = {Entropy},
VOLUME = {23},
YEAR = {2021},
NUMBER = {4},
ARTICLE-NUMBER = {462},
URL = {https://www.mdpi.com/1099-4300/23/4/462},
PubMedID = {33919678},
ISSN = {1099-4300},
DOI = {10.3390/e23040462}
}

@article{Bekemeier2025,
author = {Bekemeier, Simon and Caldeira Rêgo, Celso Ricardo and Mai, Han Lin and Saikia, Ujjal and Waseda, Osamu and Apel, Markus and Arendt, Felix and Aschemann, Alexander and Bayerlein, Bernd and Courant, Robert and Dziwis, Gordian and Fuchs, Florian and Giese, Ulrich and Junghanns, Kurt and Kamal, Mohamed and Koschmieder, Lukas and Leineweber, Sebastian and Luger, Marc and Lukas, Marco and Maas, J\"urgen and Mertens, Jana and Mieller, Bj\"orn and Overmeyer, Ludger and Pirch, Norbert and Reimann, Jan and Schr\"ock, Sebastian and Schulze, Philipp and Schuster, J\"org and Seidel, Alexander and Shchyglo, Oleg and Sierka, Marek and Silze, Frank and Stier, Simon and Tegeler, Marvin and Unger, J\"org F. and Weber, Matthias and Hickel, Tilmann and Schaarschmidt, J\"org},
title = {Advancing Digital Transformation in Material Science: The Role of Workflows Within the MaterialDigital Initiative},
journal = {Advanced Engineering Materials},
volume = {27},
number = {8},
pages = {2402149},
doi = {https://doi.org/10.1002/adem.202402149},
url = {https://advanced.onlinelibrary.wiley.com/doi/abs/10.1002/adem.202402149},
eprint = {https://advanced.onlinelibrary.wiley.com/doi/pdf/10.1002/adem.202402149},
year = {2025}
}

@article{JANSSEN2019,
title = {pyiron: An integrated development environment for computational materials science},
journal = {Computational Materials Science},
volume = {163},
pages = {24-36},
year = {2019},
issn = {0927-0256},
doi = {https://doi.org/10.1016/j.commatsci.2018.07.043},
url = {https://www.sciencedirect.com/science/article/pii/S0927025618304786},
author = {Jan Janssen and Sudarsan Surendralal and Yury Lysogorskiy and Mira Todorova and Tilmann Hickel and Ralf Drautz and J\"org Neugebauer}
}

@article{VASP,
  title = {Efficient iterative schemes for ab initio total-energy calculations using a plane-wave basis set},
  author = {Kresse, G. and Furthm\"uller, J.},
  journal = {Phys. Rev. B},
  volume = {54},
  issue = {16},
  pages = {11169--11186},
  numpages = {0},
  year = {1996},
  month = {Oct},
  publisher = {American Physical Society},
  doi = {10.1103/PhysRevB.54.11169},
  url = {https://link.aps.org/doi/10.1103/PhysRevB.54.11169}
}

@article{BOECK2011543,
title = {The object-oriented DFT program library S/PHI/nX},
journal = {Computer Physics Communications},
volume = {182},
number = {3},
pages = {543-554},
year = {2011},
issn = {0010-4655},
doi = {https://doi.org/10.1016/j.cpc.2010.09.016},
url = {https://www.sciencedirect.com/science/article/pii/S0010465510003619},
author = {S. Boeck and C. Freysoldt and A. Dick and L. Ismer and J. Neugebauer}
}

@inbook{Engelhardt,
author = {Larry Engelhardt and Christian Schr\"oder},
title = {Simulating Computationally Complex Magnetic Molecules},
booktitle = {Molecular Cluster Magnets},
chapter = {},
pages = {241-296},
doi = {10.1142/9789814322959_0006},
year = {2011},
publisher="World Scientific",
isbn = {978-981-4322-94-2},
URL = {https://www.worldscientific.com/doi/abs/10.1142/9789814322959_0006},
eprint = {https://www.worldscientific.com/doi/pdf/10.1142/9789814322959_0006}
}

@misc{Schroder_CINOLA_-_Classical_2023,
author = {Schr\"oder, Christian and Bekemeier, Simon},
month = oct,
title = {{CINOLA - Classical Spin Dynamics for Core / Shell Particles (binary distribution)}},
url = {https://github.com/s4b7r/CINOLA-binary-distribution},
version = {0.0.1},
year = {2023}
}

@misc{Bekemeier_CINOLA_pyiron_Workflow_2023,
author = {Bekemeier, Simon and Schr\"oder, Christian},
month = oct,
title = {{CINOLA pyiron Workflow}},
url = {https://github.com/s4b7r/cinola-pyiron-workflow},
version = {0.0.1},
year = {2023}
}

@article{Taake24,
    author = {Taake, Chris and Samanta, Tapas and Caron, Luana},
    title = {Field-sensitivity and reversibility of the inverse magnetocaloric effect at martensitic transformations},
    journal = {Applied Physics Letters},
    volume = {124},
    number = {5},
    pages = {052403},
    year = {2024},
    month = {01},
    issn = {0003-6951},
    doi = {10.1063/5.0185552},
    url = {https://doi.org/10.1063/5.0185552},
    eprint = {https://pubs.aip.org/aip/apl/article-pdf/doi/10.1063/5.0185552/20134303/052403_1_5.0185552.pdf},
}

@article{CARPi2017, doi = {10.21105/joss.00146}, url = {https://doi.org/10.21105/joss.00146}, year = {2017}, publisher = {The Open Journal}, volume = {2}, number = {12}, pages = {146}, author = {Nicolas CARPi and Alexander Minges and Matthieu Piel}, title = {eLabFTW: An open source laboratory notebook for research labs}, journal = {Journal of Open Source Software} }

@book{Planes,
author = {Planes, Antoni and Mañosa, L. and Saxena, A.},
year = {2005},
month = {01},
pages = {},
title = {Magnetism and Structure in Functional Materials},
volume = {79},
isbn = {978-3-540-23672-6},
journal = {Magnetism and Structure in Functional Materials},
doi = {10.1007/3-540-31631-0},
publisher={Springer}
}

@misc{mediawiki,
  title = {MediaWiki},
  url       = {https://www.mediawiki.org/wiki/Extension:OttrParser},
}

@misc{cco,
  title = {Common Core Ontology},
  url       = {https://github.com/CommonCoreOntology/CommonCoreOntologies}}

@misc{pmdco,
  title = {PMD Core Ontology},
  url       = {https://github.com/materialdigital/core-ontology},
}

@misc{qudt,
  title = {QUDT},
  url       = {https://www.qudt.org/},
}

@misc{zenodo,
  title = {DiProMag Ontology Zenodo repository},
  url       = {https://zenodo.org/records/10606363},
}

@misc{dipromagonotology,
  title = {
DiProMag Ontology},
  url       = {https://gitlab.ub.uni-bielefeld.de/semantic-computing/dipromagontology},
}

@misc{dipromag_onto,
  title = {DiProMag templates},
  url       = {https://www.dipromag.de/dipromag_onto/index.html},
}

\end{document}